\documentclass[aps,prl,twocolumn,nofootinbib,superscriptaddress
]{revtex4-2}
\usepackage{preamble}

\begin{document}

\title{Gravitational waves in hierarchical triples with lensing and relativistic orbits}

\author{Brian C. Seymour\,\orcidlink{0000-0002-7865-1052}} %
\email{seymour.brianc@gmail.com}
\affiliation{Center of Gravity, Niels Bohr Institute, Blegdamsvej 17, 2100 Copenhagen, Denmark}
\author{Fabiano Feleppa\,\orcidlink{0000-0002-3706-1021}}
\affiliation{Dipartimento di Fisica ``E.R. Caianiello'', Universit\`a di Salerno, Via Giovanni Paolo II 132, I-84084 Fisciano, Italy}
\affiliation{Istituto Nazionale di Fisica Nucleare, Sezione di Napoli, Via Cintia, 80126 Napoli, Italy}
\affiliation{Center for Fundamental Physics, School of Physical Science and Technology, ShanghaiTech University, 393 Middle Huaxia Road, Shanghai 201210, China}
\author{Luka Vujeva\,\orcidlink{0000-0001-7697-8361}}
\affiliation{Center of Gravity, Niels Bohr Institute, Blegdamsvej 17, 2100 Copenhagen, Denmark}
\author{Johan Samsing\,\orcidlink{0000-0003-0607-8741}}
\affiliation{Center of Gravity, Niels Bohr Institute, Blegdamsvej 17, 2100 Copenhagen, Denmark}

\begin{abstract}
    Binary black holes (BBH) may be formed near galactic nuclei. If such BBH systems exist, modulations to the gravitational waveform through time delays, lensing, and angular momentum precession could allow the presence of the third body to be measured in this system. In this letter, we advance the modeling for hierarchical triples by carefully including new relativistic lensing, aberration, and polarization terms.
    We show that the gravitational waves from this system can be computed exactly in the geometric optics limit by ray tracing the null geodesics and parallel transporting their polarization without a weak-field expansion. Our construction evolves the inner binary as if it were a post-Newtonian signal, and computes the environmental modulations efficiently with elliptic integrals.
    We find that this method improves the accuracy of such a triple waveform significantly compared to previous post-Newtonian approximations, and allows numerically efficient computation of the modulations. This lays the groundwork for future hierarchical triple waveform codebases, which can be used in LISA parameter estimation and detectability.
\end{abstract}

\maketitle
\begin{figure*}
    \centering
    \includegraphics[width=0.95\linewidth]{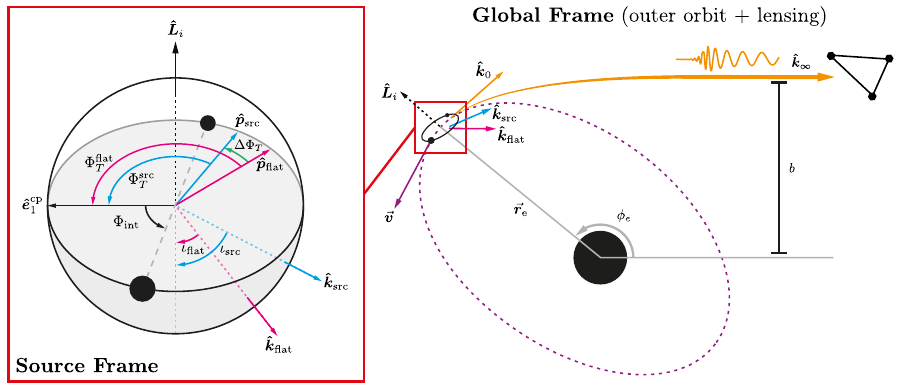}
    \caption{Geometry of the inner binary, the outer orbit, and the lensed LOS. \textit{Left:} the source frame $(\uvec e_1^\rm{cp}, \uvec e_2^\rm{cp}, \uvec L_i)$, in which the binary evolves as if isolated, with intrinsic phase $\Phi_\rm{int}$ measured from $\uvec e_1^\rm{cp}$. We compute the phase measured from the principal+ axis $\uvec p \propto \uvec k \times \uvec L_i$. In flat space, the LOS direction is $\uvec k_\rm{flat}$ with principal+ axis $\uvec p_\rm{flat}$ (pink). Lensing and aberration modify these to $\uvec k_\rm{src}$ and $\uvec p_\rm{src}$ respectively (blue). These effects are imprinted into the waveform via changes to the Thomas phase $\Phi_T$ and inclination $\iota$.
    \textit{Right:} the global frame containing the SMBH, the observer, and the binary which moves with velocity $\vec v$ (purple) along its outer orbit (dotted purple). The GWs travel along a null geodesic (orange) with lens angle $\phi_\rm{e}$ and impact parameter $b$, emitted from $\vec r_\rm{e}$ along $\uvec k_0$ and reaching the observer along $\uvec k_\infty = \uvec k_\rm{flat}$.
    Finally, the source frame direction $\uvec k_\rm{src}$ is aberrated from $\uvec k_0$ due to the binary's velocity $\vec v$. 
    }
    \label{fig:coprecessing-lensing-geometry}
\end{figure*}

\prlsec{Introduction}
A growing number of LIGO--Virgo--KAGRA (LVK) binary black hole (BBH) events, both individually \cite{LIGOScientific:2020iuh,LIGOScientific:2020ufj,Woosley:2016hmi,Belczynski:2016jno,Spera:2017fyx,LIGOScientific:2025rsn,LIGOScientific:2025brd} and at a population level \cite{Tong:2025wpz,LIGOScientific:2026ctl,Plunkett:2026pxt}, exceed the predictions of isolated stellar evolution and point to a hierarchical subpopulation. Because GW kicks tend to eject remnants from most stellar environments \cite{Merritt:2004xa,Campanelli:2007cga,Fishbach:2017dwv,Gerosa:2019zmo,Rodriguez:2019huv,Mahapatra:2021hme,Gerosa:2021mno}, repeated merger events likely require the deep potential of a galactic nucleus or massive cluster \cite{Antonini:2018auk,Yang:2019cbr,Martinez:2020lzt,Sedda:2023big,Tiwari:2026fta}. Gas-assisted mergers in active galactic nucleus (AGN) disks have also been proposed as a channel for BBHs \cite{McKernan:2012rf,Bellovary:2015ifg,Stone:2016wzz,Mckernan:2017ssq,Tagawa:2019osr,Peng:2021vzr,Vaccaro:2026nkp,Bartos:2016dgn,Yang:2020lhq,Rowan:2026mfm}. Estimates from the AGN formation channel \cite{Bartos:2016dgn,Yang:2020lhq,Rowan:2026mfm} and dynamical captures in nuclear clusters \cite{OLeary:2008myb} predict merger rate densities of $\sim 1$--$10\,\rm{Gpc^{-3}\,yr^{-1}}$, so future GW observations may help probe AGN and their role in galaxy evolution.

There has been increased interest in directly measuring the presence of the astrophysical environment in the GW signal \cite{Takatsy:2025bfk,Zwick:2025ine,Samsing:2024syt,Zwick:2024yzh,Zwick:2022dih,Zwick:2021dlg,Garg:2022nko,Hendriks:2026kys,Tagawa:2025tfd,Samsing:2025phj,Samsing:2025rxq,Zwick:2025nbg,Zwick:2026thw,Toubiana:2020drf,LISA:2022yao,Cardoso:2025npr}. The field of environmental effects probes nonvacuum signatures in spacetime such as accretion disks \cite{Speri:2022upm,Duque:2024mfw,Duque:2025yfm,Baibhav:2026iny,Dyson:2025dlj,Dyson:2026ddd} and in the context of hierarchical triples \cite{Meiron:2016ipr,Robson:2018svj,Wu:2026pwl,Yu:2020dlm,Yu:2021dqx,Laeuger:2023qyz}. In ground-based detectors from the LVK, BBH may be sensitive to the surrounding medium \cite{Barausse:2014tra,Cardoso:2019rou,CanevaSantoro:2023aol} or to line-of-sight acceleration \cite{Vijaykumar:2023tjg,Tiwari:2023cpa,Hendriks:2024zbu,Tiwari:2025aec,Roy:2026mco,Pompili:2026tdf,Ansari:2026nhb,Ansari:2026pxp,He:2026bzv,LIGOScientific:2026qni,LIGOScientific:2026fcf}, for which no evidence has been found \cite{CanevaSantoro:2023aol,Pathak:2026cik,LIGOScientific:2026fcf,Roy:2026duh}. 

Space-based detectors, however, are a natural place to search for evidence of binary extreme mass ratio inspirals (b-EMRIs) where a BBH orbits hierarchically around a supermassive black hole (SMBH). Previous works on b-EMRIs have studied how orbital dynamics modulate the GW waveform due to source motion \cite{Han:2018hby,Cisneros:2012sk,Meiron:2016ipr,Bonvin:2016qxr,Inayoshi:2017hgw,Robson:2018svj,Tamanini:2019usx,Wong:2019hsq,Zwick:2025wkt,Zwick:2025qzv}, precession of angular momentum \cite{Liu:2019tqr,Liu:2021uam}, lensing \cite{Randall:2018lnh,DOrazio:2019fbq,Leong:2024nnx,Ubach:2025dob,Ubach:2025oxr,Martens:2026mku,Li:2025xuh,Xu:2025frd,Hou:2019wdg,Oancea:2022szu,Oancea:2023hgu,Pijnenburg:2024btj,Chan:2025wgz,Saketh:2025cwf,Santos:2026mbe,Wu:2026rqm,Huang:2026hbb}, gravitational redshift \cite{Chen:2017xbi,Kuntz:2022juv,Cardoso:2026ugm}, aberration \cite{Bonvin:2022mkw,He:2022qhc,Cusin:2024git,Torres-Orjuela:2018ejx,Torres-Orjuela:2020cly,Torres-Orjuela:2020dhw} and the Lidov--Kozai mechanism \cite{Lidov:1962wjn,Kozai:1962zz,Hoang:2019kye,Fang:2019mui,Yu:2020iqj,Deme:2020ewx,Chandramouli:2021kts,Camilloni:2023rra,Camilloni:2023xvf,Zhang:2024ibf,Cocco:2025adu,Cocco:2025udb,Cocco:2026lkr}. Hierarchical triples would be powerful probes, both for the astrophysical environment of BBHs \cite{Sberna:2022qbn,Xuan:2022qkw,Saini:2025ncj,Hendriks:2024gpp} and for testing the dynamical strong-field regime of gravity \cite{Cardoso:2021vjq,Oancea:2023hgu}, just as in binary pulsar measurements \cite{Ransom:2014xla,Archibald:2018oxs,Voisin:2020lqi,Seymour:2019tir}. 

No existing b-EMRI waveform treats strong-field propagation consistently while remaining fast enough for LISA parameter estimation.
So far, the community has taken two approaches: (1) `top-down' phenomenological models, which begin from a post-Newtonian (PN) waveform and add the effects of the SMBH afterwards \cite{Meiron:2016ipr,Robson:2018svj,Tiwari:2026lvw,Yu:2020dlm,Yu:2021dqx,Laeuger:2023qyz} and (2) `bottom-up' models, which solve the Teukolsky equation numerically \cite{Yin:2024nyz,Meng:2024kug,Jiang:2024mdl,Santos:2025ass,Santos:2026lzq}.
The `top-down' models can be computed quickly for long inspirals but break down when they are too close to the SMBH. The `bottom-up' models compute the GW accurately for binaries close to the SMBH but are likely too slow for production-ready waveforms for LISA, particularly when the inner orbit is evolving.

In this Letter, we present a new `top-down' construction of triple waveforms where we use a PN-based isolated-binary waveform in its local rest frame, but we propagate the emitted GW through the SMBH's spacetime. The waveform is constructed in two steps: we first solve for the center-of-mass motion along a bound geodesic of the SMBH and parallel transport the angular momentum along it, and then we describe the lensing from the SMBH with geometric optics.
The center-of-mass motion is computed for generic $(a_o,e_o)$ using elliptic integrals in the Darwin parameterization \cite{Darwin:1959-gravityfield,Darwin:1961-gravityfieldII,Cutler:1994pb}, and the inner angular momentum is Fermi--Walker parallel transported with the Marck tetrad \cite{Marck:1983a,Marck:1983b,Kamran:1986,Akcay:2016dku}, improving on previous triple models \cite{Yu:2020dlm,Laeuger:2023qyz}. We ray trace the signal to the observer by solving the lens equation for each image with elliptic integrals \cite{Hackmann:2018fmk,Gralla:2019ceu,Ben-Salem:2022txj}. This gives us the propagation delay from proper time in the binary frame to infinity and the corresponding source magnification \cite{Frittelli:1999yf,Beloborodov:2002mr,Poutanen:2019tcd}, for each image. Finally, we relate the binary's rest frame to the observer by including ray bending during lensing \cite{Beloborodov:2002mr,Poutanen:2019tcd,Pechenick:1983a}, and aberration from the binary's velocity \cite{Bonvin:2022mkw,He:2022qhc,Cusin:2024git}. Both of these effects rotate the line-of-sight (LOS) direction in which the binary emits towards the observer, modulating the inclination, polarization angle \cite{Hou:2019wdg,Oancea:2022szu}, Thomas phase, and amplitude of each image.

We validate the model against the Teukolsky b-EMRI waveforms of Santos+2025 \cite{Santos:2025ass} in the \textit{highly strong-field regime} with an outer orbit of $\sim 10$ gravitational radii, and find mismatches of $\sim 10^{-3}$--$10^{-2}$, compared to $\sim 0.7$ for a Newtonian-lens-based model (analogous to \cite{Yu:2021dqx,Laeuger:2023qyz}). Because every new ingredient is expressed with elliptic integrals, our waveform is numerically efficient and should be fast enough for LISA parameter estimation. Throughout this letter, we work in geometric units $G = c = 1$.

\prlsec{Setup}
We consider an inner binary orbiting a non-spinning SMBH of mass $M_3$. In its local rest frame (left panel of Fig.~\ref{fig:coprecessing-lensing-geometry}), the binary evolves as an isolated binary. The PN waveform for an isolated binary is given by
\begin{equation}\label{eq:h-int}
    h_\rm{int}(\tau) \equiv \mathcal{A} \left[ A_+ \cos 2\Phi_\rm{int} - i A_\times \sin 2\Phi_\rm{int} \right]\,,
\end{equation}  
where $h_\rm{int}(\tau) \equiv h_+ - i h_\times$ is the complex time-domain strain, $\tau$ is the binary's proper time, $\mathcal{A}(\tau)$ is the amplitude, $\Phi_\rm{int}(\tau)$ is the orbital phase, and $A_{+,\times}(\iota)$ are functions of the binary's inclination $\cos\iota = - \uvec k \cdot \uvec L_i$, relative to the direction $\uvec k$ in which the wave leaves.

The waveform at infinity (right panel of Fig.~\ref{fig:coprecessing-lensing-geometry}) is a source frame PN waveform, evaluated at a delayed time, but rescaled in amplitude and viewed from a rotated direction. An observer at infinity sees a waveform composed of multiple images,
\begin{equation}
    h_\rm{o} = \sum_I h_\rm{o}^I\,,
\end{equation}
with each image being
\begin{align}\label{eq:observed-strain-setup}
    h_\rm{o}^I(t_\rm{o}) = |\mu_I|^{1/2} \frac{f_{\rm o}^I }{f_{\rm e}} \, h_\rm{int}^I(\tau)\Big|_{\tau=\tau_\rm{e}^I(t_\rm{o})}\,.
\end{align}
The amplitude of the GW strain is rescaled by the lensing magnification $|\mu_I|^{1/2}$ due to null rays focusing \cite{Beloborodov:2002mr}, and also by the Doppler magnification $f_{\rm o}^I /f_{\rm e}$, which is the ratio of observed to emitted frequencies \cite{Cusin:2024git}. Additionally, $\tau_\rm{e}^I(t_\rm{o})$ is the proper time at which image $I$ was emitted, and $h_\rm{int}^I(\tau)$ is the waveform from Eq.~\eqref{eq:h-int} evaluated along the emission direction of image $I$ in the local rest frame of the binary.

The GW observed at infinity experiences time delays between the source and observer. When the GW is emitted at proper time $\tau_\rm{e}$, this corresponds to a Schwarzschild time $t_\rm{e}$ at which the binary is at $\vec r_{\rm e}$ in its outer orbit. The observer at infinity sees image $I$ arrive at $t_{\rm o} = t_\rm{e} + \Delta_I(t_\rm{e})$, where $\Delta_I$ is the propagation delay analogous to the Rømer time delay. These relations can be inverted to compute $\tau_{\rm e}^I(t_\rm{o})$. The frequency ratio is thus the ratio of the rates between emitter's and observer's clocks\footnote{Equation~\eqref{eq:clock-chain} is equivalent to the well-known result for a Schwarzschild spacetime with a timelike Killing vector, where $f_{\rm o}^I/f_{\rm e} = \left( u_\rm{o} \cdot k_I \right)_\rm{o}/\left( u_\rm{e} \cdot k_I \right)_\rm{e}$ for a ray $k_I^\mu$ where $u_\rm{e}^\mu$ and $u_\rm{o}^\mu$ are the binary's 4-velocity at emission and the observer's 4-velocity at infinity respectively.}
\begin{align}\label{eq:clock-chain}
    \frac{f_{\rm o}^I }{f_{\rm e}} = \frac{d\tau_\rm{e}^I}{dt_o} = \frac{d\tau_\rm{e}}{dt_\rm{e}}\,\frac{dt_\rm{e}^I}{dt_o} \,.
\end{align}
The first factor contains the gravitational redshift and time dilation of the binary clock, and the second is the Doppler shift from propagation of an exact null geodesic. 

Precession, lensing, and aberration modulate the GW by changing the viewing direction relative to the inner binary. Precession of the inner angular momentum $\uvec L_i$ about the orbital angular momentum $\uvec J$ in triples has been studied \cite{Liu:2019tqr,Yu:2020dlm} using precession results in PN waveforms \cite{Apostolatos:1994mx}. In flat space, this is understood as the direction of $\uvec L_i$ changing with respect to the fixed LOS $\uvec k_\rm{flat}$. Lensing and aberration are instead \textit{rotations to the LOS orientation} $\uvec k_\rm{src}^I$, which modulates the waveform that reaches infinity. In this manuscript, we show that flat-space precession can be combined with lensing and aberration, and these orientation effects can be expressed as
\begin{equation}
    h_\rm{int}^I(\tau) \equiv e^{-2i \psi_I } \mathcal{A} \left[ A_+^I \cos 2\Phi_I - i A_\times^I \sin 2\Phi_I \right]\,,
\end{equation}
where we define
\begin{align}\label{eq:additional-pol-thomas-phase}
    \Phi_I &= \Phi_\rm{int} + \Phi_T^\rm{prec} + \Delta\Phi_{T,I} + \frac{\pi n_I}{4} \,, \\
    \psi_I &= \psi^\rm{prec} + \Delta \psi_I \,, \qquad A_{+,\times}^I = A_{+,\times}(\iota_I)\nn \,,
\end{align}
where $n_I$ is the Morse index of the image ($n_1 = 0$, $n_2 = 1$) \cite{Takahashi:2003ix}, and $\cos\iota_I = - \uvec k_\rm{src}^I \cdot \uvec L_i$. The quantities $(\Phi_T^\rm{prec},\psi^\rm{prec})$ are the Thomas phase and polarization phase due to precession in flat space \cite{Apostolatos:1994mx,Liu:2019tqr,Yu:2020dlm}. For the first time, we systematically compute the new terms $(\iota_I, \Delta \psi_I, \Delta \Phi_{T,I})$ from lensing and aberration. 

In the rest of this Letter, we will compute each ingredient of Eq.~\eqref{eq:observed-strain-setup} and construct the waveform in three steps. First, we evolve the outer orbit to get $\vec r_\rm{e}(t_\rm{e})$, $\tau_\rm{e}(t_\rm{e})$, and $\uvec L_i$. Second, we ray trace the geodesic to get $\Delta_I(t_\rm{e})$ and $\mu_I$. Finally, we compute modifications from the rotated LOS due to bending and aberration $(\iota_I,\Delta \psi_I,\Delta \Phi_{T,I})$.

\prlsec{Outer orbital motion}
We compute the trajectory of the binary's center of mass by solving the geodesic equation for a test mass. The location of the binary is given by $x^\mu(\tau)$, and the geodesic equation is $u^\mu \nabla_\mu u^\nu = 0$ where $u^\mu = dx^\mu/d\tau$.
We follow the framework laid out in Darwin's work \cite{Darwin:1959-gravityfield}, which allows for an efficient elliptic-integral representation of the system. This is solved efficiently by introducing the relativistic anomaly $\chi$ where we compute radius, azimuthal angle, and Schwarzschild emission time $(r(\chi),\varphi(\chi),t_\rm{e}(\chi))$. Bound massive geodesics are fully characterized by the semimajor axis and eccentricity $(a_o,e_o)$.

The motion of the binary's angular momentum is also computed by parallel transporting it. We will use $S^\mu$ to denote the 4-vector of the inner angular momentum. The goal for this is to solve the parallel transport equation $D S^\nu/d\tau = u^\mu \nabla_\mu S^\nu = 0$. This does not need to be numerically integrated in practice because we can construct the Marck tetrad \cite{Marck:1983a,Marck:1983b}, which is trivially parallel transported.
In practice, we express the parallel transport of the inner angular momentum in terms of the relativistic anomaly as laid out in Ref.~\cite{Akcay:2016dku}.
The cumulative precession phase is given by $\alpha(t)$ which can be generically computed for any $(a_o,e_o)$ values. This approach improves on previous triple codes which used PN approximations for the inner angular momentum precession \cite{Yu:2020dlm,Laeuger:2023qyz}. A detailed explanation of the motion and precession is given in the Supplemental Material~\cite{supp} (App.~\ref{app:motion-precession}).

\prlsec{Geometric optics approach to lensing} 
We compute the null rays by solving the geodesic equation for rays that connect the point $x^\mu(\tau)$ to the observer at infinity, which corresponds to the geometric optics limit in lensing. Due to the symmetries of the Schwarzschild SMBH, solving the geodesic equation $k^\mu \nabla_\mu k^\nu = 0$ for a photon ray $k^\mu(\lambda)$ can be reduced to finding the value of the impact parameter $b\equiv L/E$ so that the null geodesic passes through a particular point in spacetime. In practice, we use the elliptic integral solutions for the angular sweep $\Delta \phi(b)$ \cite{Darwin:1959-gravityfield,Chandrasekhar:1985kt,Hackmann:2018fmk} to solve the lens equation $\Delta \phi(b) = \phi_\rm{e}, 2\pi - \phi_\rm{e}$ for the primary and secondary images, respectively, which gives the impact parameter $b_I$ of image $I$.

With this impact parameter in hand, the results of Hackmann \cite{Hackmann:2018fmk} give closed-form elliptic expressions for the time delay function
\begin{equation}\label{eq:time-delay-full-main}
    \Delta_I(t_\rm{e}) \equiv \Delta t_I(t_\rm{e}) - \Delta t_1(t_\rm{ref}) \,,
\end{equation}
where $\Delta t_I(t_\rm{e}) \equiv \Delta t(r_\rm{e}(t_\rm{e}),b_I(t_\rm{e}))$ and $\Delta t_1(t_\rm{ref})$ is a reference time delay for the first image. There is a closed-form expression for $\Delta t_I(t_\rm{e}) $ in terms of a sum of incomplete elliptic integrals, which is explicitly given in the Supplemental Material. The expression in Eq.~\eqref{eq:time-delay-full-main} allows us to efficiently compute the time delay between emission and observation $t_o^I(t_\rm{e}) = t_\rm{e} + \Delta_I(t_\rm{e})$ which is used in Eq.~\eqref{eq:clock-chain}.
Finally, the magnification in the geometric optics limit is given by \cite{Beloborodov:2002mr,Poutanen:2019tcd}
\begin{equation}
    \mu(r_\rm{e},b) = \frac{b}{r_\rm{e}^2 |\cos \beta_\rm{e}| \left|\sin \phi_\rm{e}\right|\left|\partial_b \Delta \phi(b)\right|}\,,
\end{equation}
where $\cos \beta_\rm{e} \equiv \uvec k_0^I \cdot \uvec r_\rm{e}$ is the degree to which the GW was tangentially emitted, and $\uvec k_0^I$ is the null geodesic direction in the static frame.

\prlsec{Aberration and polarization}
Since the binary is moving through spacetime, the GW is modulated due to bending during lensing and aberration. The magnification factor from aberration is already well understood from \cite{Cusin:2024git} (cf.~Eq.~\eqref{eq:observed-strain-setup}). Here we focus on the remaining effects from rotations that the observed ray leaves in the source frame. As we show in Fig.~\ref{fig:coprecessing-lensing-geometry}, lensing bends $\uvec k_0^I$ into the asymptotic direction $\uvec k_\infty$, while aberration tilts the direction in which the ray leaves the binary's source frame, $\uvec k_\rm{src}^I$, away from the static-frame direction $\uvec k_0^I$. These two respective angles, $\cos\delta_I = \uvec k_0^I\cdot\uvec k_\infty$ and $\cos\epsilon_I = \uvec k_\rm{src}^I\cdot\uvec k_0^I$ are given by
\begin{align}\label{eq:lens-aber-angle}
    \delta_I &\simeq \frac{4M_3}{b_I}\sin^2\frac{\phi_\rm{e}}{2} \,,&
    \epsilon_I &\simeq v \sin\theta_0^I \,, 
\end{align}
to leading order in $v$ and $M_3/b_I$, where $\cos\theta_0^I \equiv \vec v\cdot\uvec k_0^I / \vert \vec v\vert$. For the primary image, the deflection is $\delta_1 = \phi_\rm{e} - \beta_\rm{e}$ exactly.  As we prove in the Supplemental Material, the observed direction in the binary's source frame that the ray comes from is
\begin{align}\label{eq:rotated-los-src}
    \uvec k_\rm{src}^I &= \mat R_\rm{ab}^I\mat R_\rm{lens}^I \uvec k_\rm{flat} \,, 
\end{align}
where $\mat R_\rm{lens}^I$ and $\mat R_\rm{ab}^I$ are rotation matrices which compactly express the transformation for lensing and aberration by angles $\delta_I,\epsilon_I$ respectively. Additionally, the source frame principal+ axis is recomputed from the rotated LOS via $\uvec p_\rm{src}^I \propto \uvec k_\rm{src}^I \times \uvec L_i $, compared to the case without lensing and aberration, $\uvec p_\rm{flat} \propto \uvec k_\rm{flat} \times \uvec L_i $. Both lie in the binary's orbital plane, and the rotation of $\uvec p_\rm{flat}$ into $\uvec p_\rm{src}^I$ about $\uvec L_i$ is the Thomas phase correction $\Delta \Phi_{T,I}$, as shown in Fig.~\ref{fig:coprecessing-lensing-geometry}. Thus, $\uvec k_\rm{flat}$ is fixed in time, and all the time-dependent modulations to the waveform are encoded through the LOS rotations from precession, lensing, and aberration. 

Expressing lensing and aberration for the propagation and principal+ axis with rotation matrices in Eq.~\eqref{eq:rotated-los-src} allows us to concretely identify modulations to the inclination, Thomas phase, and polarization angle in Eq.~\eqref{eq:additional-pol-thomas-phase}. The inclination changes as $\cos \iota_I = -\uvec k^I_{\rm{src}}\cdot\uvec L_i$ while the rotation of the polarization phase $\Delta \psi_I = \Delta \psi^\rm{lens}_I + \Delta \psi^\rm{ab}_I$ and Thomas phase $\Delta \Phi_{T,I} = \Delta \Phi^\rm{lens}_{T,I} + \Delta \Phi^\rm{ab}_{T,I}$ can be broken down into additional rotation terms explicitly. To compute the polarization rotation $\Delta\psi_I$, we carry $\uvec p_\rm{src}$ to the observer by undoing the aberration and then parallel transporting it along the geodesic and comparing it to the flat space case $\uvec p_\rm{flat}$. Ultimately, \textit{aberration and lensing LOS modulations can be systematically computed with rotation matrices} in this fully relativistic b-EMRI case.

\begin{figure}[h]
    \centering
    \includegraphics{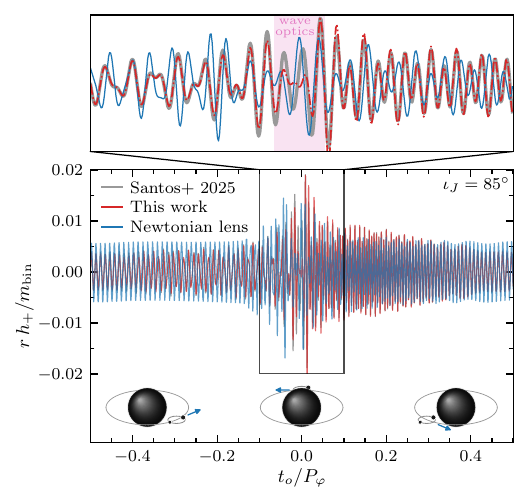}
    \caption{The waveform generated by our triple code (red) compared to the Teukolsky-based b-EMRI data (grey) \cite{Santos:2025ass} and a Newtonian lens waveform (blue). We plot this for a highly aligned system which is nearly edge-on with $a_o = 10 M_3$, $\iota_J = 85^\circ$, and $M_3 \omega_\rm{int} = 2.16$. One can see that our triple waveform nearly perfectly matches the b-EMRI waveform as long as the geometric optics regime is valid. When the system is nearly aligned, geometric optics breaks down due to diffraction once the time delay between images is $\Delta t_l \lesssim \lambda_\rm{gw}$ \cite{Nakamura:1999uwi,Takahashi:2003ix} (pink shaded region, inset). 
    }
    \label{fig:comparison-inc-85}
\end{figure}

\prlsec{Results}
With this framework, we can compute how a GW is modulated when its source orbits an SMBH. In Fig.~\ref{fig:comparison-inc-85}, we plot $h_+$ for a circular binary with outer orbital radius of $a_o = 10 M_3$ and an inclination of $\iota_J = 85^\circ$, so the system is nearly edge-on (where $\cos \iota_J \equiv -\uvec J \cdot \uvec k_\infty$). The binary's internal angular frequency is $M_3 \omega_\rm{int} = 2.16$. We show the dimensionless strain $r h_+ / m_\rm{bin}$ for this system, where the inner angular momentum is aligned with the orbital angular momentum ($\uvec L_i \parallel \uvec J$). On the x-axis, we plot the observed time divided by the orbital period where maximal alignment is at $t_\rm{o} = 0$. One can see that the strain is significantly lensed there, and a beating pattern emerges due to interfering GW images, along with the aberration and time delays causing the phase and amplitude to modulate. We demonstrate that our ray-traced PN waveform has remarkable agreement with the b-EMRI Teukolsky data (grey) from \cite{Santos:2025ass}, which is the only available model which contains all strong-field effects. The only region of disagreement occurs when geometric optics breaks down due to diffractive contributions in the lensed signal (when $\Delta t_l \lesssim \lambda_\rm{gw}$ as shown in the pink shaded region). Our waveform agrees with the Teukolsky-based triple waveform, and generation is computationally efficient ($\sim 0.01$ seconds for Fig.~\ref{fig:comparison-inc-85}).

\begin{figure}[h]
    \centering
    \includegraphics{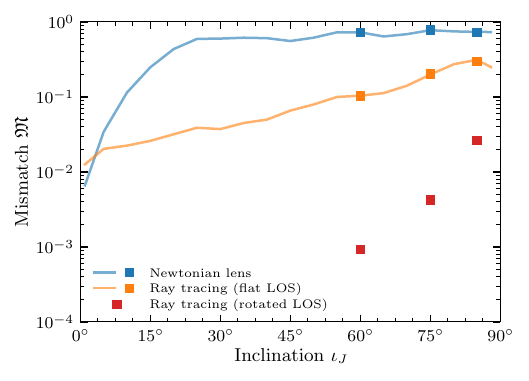}
    \caption{Mismatch of hierarchical triple waveforms as a function of inclination angle for a circular orbit with $a_o = 10M_3$. We show the Newtonian lens model (blue), the simple ray-traced model without lensing and aberration rotations (orange), and the full hierarchical triple model with LOS rotations from bending and aberration (red). In squares, we compare the accuracy of these models to the Teukolsky result \cite{Santos:2025ass}, while the lines show mismatch of the approximant waveform compared to our full one. This demonstrates our full lens model is able to improve the mismatch by around two orders of magnitude for the case of the Newtonian lens model (which is analogous to previous PN triple models \cite{Yu:2021dqx,Laeuger:2023qyz}). At high inclinations the mismatch is higher due to lensing being more significant when it is nearly edge on.
    }
    \label{fig:mismatch-vs-theta}
\end{figure}

In Fig.~\ref{fig:mismatch-vs-theta}, we compare the accuracy of waveforms that are generated with different methods as a function of the inclination $\iota_J$. The mismatch is shown on the y-axis, and squares compare the accuracy of different waveforms to the ones from \cite{Santos:2025ass} for inclinations of $\iota_J = 60^\circ, 75^\circ, 85^\circ$. First, we compute the waveform in the case of having Newtonian estimates of lensing and precession, similarly to how it was done in \cite{Yu:2021dqx} (blue), which does not use any ray tracing. One can see the waveform with ray-traced time delays (orange) but no adjustments from bending and aberration performs better than the nonrelativistic one. Finally, we show the mismatch for different observation directions for the geometric optics waveform procedure computed in this paper (red). One can see that it achieves a significantly better mismatch due to properly accounting for the relativistic effects. The solid lines in this plot compare the approximate waveforms (blue and orange) to the waveform computed in this paper (red) to give a continuous estimate of the mismatch. In general, the accuracy of all of the waveforms degrades as we approach an edge-on system due to the aforementioned diffraction/wave optics features when the system is aligned. This figure shows that our model improves upon the previous waveforms by around two orders of magnitude compared to previous PN waveforms. The agreement between our waveform here and the Teukolsky estimates \cite{Santos:2025ass} should be the weakest when the binary is so close to the SMBH $(a_o = 10 M_3)$, so this validation test shows that this method will perform even better at larger separations. We explain our method for computing the mismatch and give additional validation tests in the Supplemental Material.

\prlsec{Conclusion}
In this manuscript, we have laid out our results for computing modifications to the GW from lensing and aberration in hierarchical triples. By carefully combining ray-traced geometric optics, exact outer orbit evolution with the Marck tetrad, and modeling the lensing/aberration between the local binary's frame and the global one, we have developed a fast, accurate method for computing hierarchical triple waveforms. We stress that while we have shown the efficacy of this method in the circular, aligned inner angular momentum case as a proof of principle, our method works for \textit{arbitrary} eccentric outer orbits and inner angular momentum configurations.
Because the inner binary enters only through its source-frame waveform, \textit{generic inner waveforms} can be flexibly inserted with our framework. This hierarchical triple waveform is both accurate in the strong-field regime, and likely fast enough for parameter estimation and detectability studies in LISA (cf.~Fig.~\ref{fig:comparison-inc-85} took $\sim 0.01$ seconds to generate). In an upcoming paper, we will release this as public Python code for an evolving inner binary.

In the future, it is important to include tidal effects such as the Lidov--Kozai mechanism, and possible accretion disk matter effects. It may be possible to generalize this to the case of less asymmetric mass ratios and to the wave optics case \cite{Vujeva:2026rqh,Santos:2026mbe}. The self-force community is interested in b-EMRI waveforms, so the results could be useful in this intersection, particularly for the adiabatic early-time inspiral. Finally, understanding detectability of such systems should be investigated further, and it is possible that hierarchical triples can offer new capabilities for testing GR in the LISA era.

\prlsec{Acknowledgments}
We thank Ana Carvalho for drawing Fig.~\ref{fig:coprecessing-lensing-geometry}, and João Sieiro dos Santos for his Teukolsky b-EMRI data. We thank Hang Yu, Lorenz Zwick, Maarten van de Meent, Ariadna Ribes Metidieri, Giada Caneva Santoro, Vitor Cardoso, Connar Rowan, Chris Tiede, Andrew Laeuger, and Colin Weller for their discussions. Claude Code (Opus 5.5) was used to assist with code development and debugging, consistency checks, and figure preparation. All outputs have been independently verified by the authors, who take full responsibility for the correctness of this manuscript. 

BCS is supported by the European Union's Horizon Europe research and innovation programme under the Marie Skłodowska-Curie grant agreement No.~101277678 (TripleGW). FF gratefully acknowledges the hospitality of the Center of Gravity at the Niels Bohr Institute, where part of this work was carried out. FF acknowledges support from the Istituto Nazionale di Fisica Nucleare (INFN) through the Commissione Scientifica Nazionale 4 (CSN4) Iniziativa Specifica ``Quantum Universe'' (QGSKY). JS is supported by the Villum Fonden grant No. 29466, and by the ERC Starting Grant no. 101043143 Black HoleMergs. The Center of Gravity is a Center of Excellence funded by the Danish National Research Foundation under grant No.~184. This project has received funding from the European Union's Horizon 2020 research and innovation programme under the Marie Skłodowska-Curie grant agreement No 101299389 and No 101131233. This work is supported by Simons Foundation International \cite{sfi} and the Simons Foundation \cite{sf} through Simons Foundation grant SFI-MPS-BH-00012593-11.

\section*{Supplemental Material}

\setcounter{secnumdepth}{3}
\setcounter{equation}{0}
\setcounter{figure}{0}
\renewcommand{\thesubsection}{S.\arabic{subsection}}
\renewcommand{\thesubsubsection}{\alph{subsubsection}}
\renewcommand{\theequation}{S.\arabic{equation}}
\renewcommand{\thefigure}{S.\arabic{figure}}
\renewcommand{\theHequation}{S.\arabic{equation}}
\renewcommand{\theHfigure}{S.\arabic{figure}}

In this Supplemental Material, we will give our approach for computing the waveform internally. In Sec.~\ref{app:motion-precession}, we explain how the binary orbits the SMBH and how its inner angular momentum precesses. In Sec.~\ref{app:geometric-optics-lensing}, we give the procedure for ray tracing the GW in a Schwarzschild spacetime using elliptic integrals. In Sec.~\ref{app:matching-source-frame}, we explain how modulations from lensing and aberration rotate the LOS of the sourced GW. Finally, in Sec.~\ref{app:validation-mismatch}, we explain how the mismatch of our waveforms is computed and benchmark the performance of the modifications against some possible triple waveform approximants.

Throughout this Supplemental Material, we use the following notation for the Jacobi incomplete elliptic integrals:
\begin{subequations}\label{eq:jacobi-elliptic-integrals}
    \begin{align}
        F(x, k) & =\int_0^x \frac{d x'}{\sqrt{\left(1-x'^2\right)\left(1-k^2 x'^2\right)}}\,, \label{eq:incomplete-elliptic-1st}\\
        E(x, k) & =\int_0^x \frac{\sqrt{1-k^2 x'^2}}{\sqrt{1 - x'^2}} d x'\,, \\
        \Pi(x, c, k) & =\int_0^x \frac{d x'}{\left(1-c x'^2\right) \sqrt{\left(1-x'^2\right)\left(1-k^2 x'^2\right)}}\,,
    \end{align}
\end{subequations}
with modulus $k$ and characteristic $c$. 

\subsection{Inner binary motion and precession}\label{app:motion-precession}

In this section, we explain our approach for modeling the inner binary's outer orbital motion and the precession of its angular momentum.

\subsubsection{Bound orbits in the Darwin parameterization}

In this manuscript, we compute the bound orbit geodesics in Schwarzschild with generic semimajor axis and eccentricity. The framework for efficiently doing this was introduced by Darwin and is called the Darwin parameterization \cite{Darwin:1959-gravityfield,Darwin:1961-gravityfieldII,Cutler:1994pb} which puts the expressions in a computationally efficient elliptic integral form. Following Ref.~\cite{Cutler:1994pb}, we express the radial and azimuthal motion in terms of the \textit{relativistic anomaly} $\chi$ as
\begin{subequations}\label{eq:r-phi-chi-darwin}
    \begin{align}
        r(\chi) &=\frac{p M_3}{1+e \cos \chi} \,,\\
        \varphi(\chi) &=p^{1 / 2} \int_0^\chi \frac{d \chi^{\prime}}{\left(p-6-2 e \cos \chi^{\prime}\right)^{1 / 2}}\,, \label{eq:phi-eq-chi}
    \end{align}
\end{subequations}
where $r,\varphi$ are the distance and azimuthal orbital phase for the bound geodesic, and $p,e$ are the \textit{semi-latus rectum} and \textit{eccentricity}, respectively. These are related to the conserved quantities of the bound orbit $\tilde{E},\tilde{L}$ which are the dimensionless energy and angular momentum 
\begin{subequations}
    \begin{align}
        \tilde{E}^2 &=\frac{(p-2-2 e)(p-2+2 e)}{p\left(p-3-e^2\right)}\,, \\
        \tilde{L}^2 &=\frac{p^2 M_3^2}{p-3-e^2} \,.
    \end{align}
\end{subequations}
The relationship between the time and relativistic anomaly is given by the integral
\begin{align}\label{eq:t-eq-chi-darwin}
    t(\chi) =&\;  p^2 M_3(p-2-2 e)^{1 / 2}(p-2+2 e)^{1 / 2} \nn \\
    & \times \int_0^\chi d \chi^{\prime}\left(p-2-2 e \cos \chi^{\prime}\right)^{-1}\left(1+e \cos \chi^{\prime}\right)^{-2}\nn \\
    & \times\left(p-6-2 e \cos \chi^{\prime}\right)^{-1 / 2}\,.
\end{align}
The radial period is given by $P_r = t(2\pi)$ and the pericenter precession $\Delta \varphi=\varphi(2 \pi)$. The pericenter advance is given by 
\begin{equation}\label{eq:delta-azimuthal-bound}
    \Delta \varphi=4\left(\frac{p}{p-6+2 e}\right)^{1 / 2} K\left(\sqrt{\frac{4 e}{p-6+2 e}}\right)\,,
\end{equation}
where $K(k) \equiv F(1,k)$ is a complete elliptic integral of the first kind matching the conventions of Eq.~\eqref{eq:jacobi-elliptic-integrals} \cite{Gradshteyn:1943cpj}.

The radial and azimuthal frequencies are thus given by \cite{Cutler:1994pb}
\begin{subequations}\label{eq:radial-azimuthal-frequencies}
    \begin{align}
        \Omega_r &=\frac{2 \pi}{P_r}\,, \\
        \Omega_\varphi &=\frac{\Delta \varphi}{2 \pi} \Omega_r \,.
    \end{align}
\end{subequations}
In principle, one has to compute $(\varphi(\chi),r(\chi),t(\chi))$ for a long $\chi$ which would be time-consuming. However, \textit{all this infrastructure was written down because one can show that} $r(t)$ and $\varphi(t)-\Omega_\varphi t$ are both periodic with $P_r$. This single fact allows us to compute the integrals in Eq.~\eqref{eq:phi-eq-chi} and Eq.~\eqref{eq:t-eq-chi-darwin} for a single radial cycle, and exploit the cyclic nature of $\varphi-\Omega_\varphi t$ to compute solutions valid for all $t$.

\subsubsection{Precession with Marck tetrad}

The binary's inner angular momentum precesses around the outer orbital angular momentum in GR \cite{Liu:2019tqr,Yu:2020dlm}, which is known as the de Sitter effect in the weak-field limit \cite{deSitter:1916zz,Barker:1975ae}. For generic orbits in GR, this can be computed by parallel transporting the angular momentum. In some previous `top-down' triple works, the precession of the binary's angular momentum was computed with PN approximations for outer orbits which are both circular \cite{Yu:2020dlm} and eccentric \cite{Laeuger:2023qyz}. This is introduced so that the precession matches what is computed in full GR as done in the binary EMRI works \cite{Santos:2025ass}. In this manuscript, we compute this in a relativistically consistent manner by constructing the Marck tetrad \cite{Marck:1983a,Marck:1983b,Kamran:1986} where the inner angular momentum is trivially transported. In this manuscript, we will focus on elliptic integral schemes in the Schwarzschild case \cite{Akcay:2016dku,Mathews:2025nyb}, though closed-form elliptic integral formulations exist in Kerr using Mino time \cite{vandeMeent:2019cam}. 

In this section, we will use $S^\mu$ to denote the 4-vector of the angular momentum. The goal for this is to solve the parallel transport equation 
\begin{equation}
    \frac{D S^\nu}{d\tau} = u^\mu \nabla_\mu S^\nu = 0 \,,
\end{equation}
along a geodesic which is parameterized by proper time $\tau$. This can be done by constructing the Marck tetrad $e_{\hat a}^\mu(\tau)$ where $g_{\mu\nu} e_{\hat{a}}{ }^\mu(\tau)e_{\hat{b}}{ }^\nu(\tau) = \eta_{\hat a \hat b}$ where $\eta_{\hat a \hat b} = \text{diag}\left[ -1,1,1,1 \right]$ and $e_{\hat{0}}{ }^\mu(\tau) = u^\mu(\tau)$. In Schwarzschild, the direction of $S^\mu$ which is parallel to the orbital angular momentum is conserved, and the other two in-plane components precess. The radial and azimuthal directions of the binary's frame are rotated by an angle $\Psi(\tau)$ which is given by \cite{Akcay:2016dku}
\begin{equation}\label{eq:tetrad-phase-deriv-proper-time}
    \frac{d \Psi}{d \tau}=\omega_{13}= \tilde E \tilde L /\left(r^2 + \tilde L^2\right) \,,
\end{equation}
where $\tau$ is the proper time along the worldline of the binary. Relative to a distant frame, the in-plane component of $S^\mu$ thus rotates by $\varphi - \Psi$. We can explicitly express this in terms of the relativistic anomaly
\begin{equation}\label{eq:tetrad-phase-deriv-rel-anomaly}
    \frac{d \Psi}{d \chi} =\sqrt{\frac{p-3-e^2}{p-6-2 e \cos \chi}} \frac{\sqrt{(p-2-2 e)(p-2+2 e)}}{\left(p-2+2 e \cos \chi-e^2 \sin ^2 \chi\right)} \,,
\end{equation}
where we used the chain rule between the rate at which the proper time elapses relative to the relativistic anomaly
\begin{equation}\label{eq:proper-time-derivative-chi}
    \frac{d \tau}{d \chi} = \frac{M_3 p^{3 / 2}}{(1+e \cos \chi)^2} \sqrt{\frac{p-3-e^2}{p-6-2 e \cos \chi}} \,.
\end{equation}
where Eq.~\eqref{eq:tetrad-phase-deriv-rel-anomaly} and Eq.~\eqref{eq:proper-time-derivative-chi} match Eq.~(11) of \cite{Akcay:2016dku} and Eq.~(2.6) of \cite{Akcay:2015pza} respectively. Since this expression also only depends on trigonometric functions of the relativistic anomaly, one can exploit this periodicity to show $\Psi(t) - \Omega_\Psi t$ is periodic with the radial period $P_r$ where we define 
\begin{align}
    \Omega_\Psi &\equiv \frac{\Delta \Psi}{P_r}\,,
\end{align}
where $\Delta \Psi\equiv \Psi(2\pi)$ \cite{Akcay:2016dku}. Thus, the \textit{same efficient numerical interpolation scheme works for the frame precession} of a generic eccentric bound orbit as for the phase in Eq.~\eqref{eq:delta-azimuthal-bound}. The same interpolation scheme gives $\tau(t)$ for all $t$. This works because Eq.~\eqref{eq:tetrad-phase-deriv-rel-anomaly} matches the form in Eq.~\eqref{eq:delta-azimuthal-bound} after accounting for different precession rates. Note that the frame precesses once per orbit, so the physical amount that the angular momentum precesses is given by 
\begin{equation}
    \alpha(t) = \varphi(\chi(t)) - \Psi(\chi(t))\,,
\end{equation}
whose secular growth rate per period matches the de Sitter precession in the weak-field limit.

\subsection{Geometric optics approach to lensing}\label{app:geometric-optics-lensing}

\subsubsection{Ray tracing}

Suppose that we have a GW emitted at location $(r_\rm{e},\phi_\rm{e})$, and the observer is located at infinity, the question we will discuss is how the geodesic can be identified which connects these two points. For a nonspinning BH, null geodesics are planar and have energy $E$ and total angular momentum $L$ conserved. The geodesic equation thus reduces to
\begin{subequations}\label{eq:null-geodesic-b}
    \begin{align}
        \frac{d t}{d \lambda} &=\frac{1}{1-2 M_3 / r}\,, \\
        \left(\frac{d r}{d \lambda}\right)^2 &=1-\frac{b^2}{r^2}\left(1-\frac{2 M_3}{r}\right)\,, \\
        \frac{d \phi}{d \lambda}&=\frac{b}{r^2} \,,
    \end{align}
\end{subequations}
where these equations only depend on the ratio of angular momentum to energy which is called the impact parameter $b = L/E$\footnote{This is possible because one can arbitrarily rescale the affine parameter $\lambda \rightarrow E \lambda$ since $k_\mu k^\mu = 0$.}. We want to integrate Eqs.~\eqref{eq:null-geodesic-b} to find the values of $b$ which connect the locations for the \textit{binary emission} location $(r_\rm{e},\phi_\rm{e})$ to the \textit{observer at infinity} $(r_\rm{o},\phi_\rm{o})$ where we henceforth consider $r_\rm{o} \rightarrow \infty$ and $\phi_\rm{o} = 0$.

If our binary is located at position $\uvec r_\rm{e}$, and the observer is located on axis $\uvec N_\infty$, this corresponds to solving for $b$ such that 
\begin{equation}\label{eq:lens-equation}
    \cos \Delta \phi(b) = - \uvec N_\infty \cdot \uvec r_\rm{e}
\end{equation}
where $\Delta \phi_1 = \phi_\rm{e}$ and $\Delta \phi_2 = 2\pi - \phi_\rm{e}$ for the primary and secondary images respectively. Explicitly, Eq.~\eqref{eq:null-geodesic-b} gives the azimuthal phase accumulation as 
\begin{equation}\label{eq:deltaphi-integral}
    |\Delta \phi(b)|\equiv\int_{\lambda_\rm{e}}^{\lambda_o} d\lambda \frac{|b|}{r(\lambda)^2} \,,
\end{equation}
where $\lambda_\rm{e}$ and $\lambda_\rm{o}$ are the affine parameters at emission and the observer's location respectively. Therefore, we just need to root find the values of $(b_1,b_2)$ which solve the lens equation in Eq.~\eqref{eq:lens-equation}.

Let us now convert the expressions in Eq.~\eqref{eq:null-geodesic-b} into ones that are suitable for evaluation in this paper using elliptic integrals. This was originally done by Darwin \cite{Darwin:1959-gravityfield} and notably discussed in \S 20 of \cite{Chandrasekhar:1985kt} for Schwarzschild. This section of the paper will follow the treatment of Hackmann \cite{Hackmann:2018fmk} which computed both the angular integral and time delay integral in terms of elliptic integrals.
The key step is to factorize the radial geodesic equation 
\begin{align}\label{eq:geodesic-equations-radial-roots}
    \left(\frac{d r}{d \lambda}\right)^2 &= \frac{b^2}{r^4} R(r) \\
    R(r) & =\frac{1}{b^2}\left(r-r_4\right)\left(r-r_3\right)\left(r-r_2\right)\left(r-r_1\right)\,, \nn
\end{align}
where $r_1<r_2=0$. The geodesics are `scattered rays' when $b>b_\rm{c} = \sqrt{27}M_3$ where $r_1<r_2<r_3<r_4$ and `captured rays' when $b<b_\rm{c}$ where $r_3,r_4 \in \mathbb{C}$. Scattered rays are unbound while the captured rays start at infinity and terminate by falling into the BH. We compute the roots using results from François Viète's method for the depressed cubic equation [explicitly using forms in \cite{Gates:2020sdh}]. For scattered rays, $r_4$ is the point at which the null ray most closely approaches the BH. If we change variables from $\lambda$ to $r$ and use the root structure of Eq.~\eqref{eq:geodesic-equations-radial-roots}, we can write the deflection of the geodesic as
\begin{equation}\label{eq:phie-integral}
    \phi_\rm{e}(r_\rm{e},b)=I_\phi(\infty,b) \pm I_\phi(r_\rm{e},b) \,,
\end{equation}
where the $+$ term corresponds to the `inward' direction and the `$-$' is the `outward' term and we define 
\begin{align}\label{eq:I_phi-integral}
    I_\phi(r,b) &\equiv \int_{r_4}^{r} \frac{d r}{\sqrt{R(r)}} \,,\nn\\
    &= \frac{2 b}{\sqrt{r_4\left(r_3-r_1\right)}} F(x;k) \,,
\end{align}
where $F$ is an incomplete elliptic integral of the first kind defined in Eq.~\eqref{eq:incomplete-elliptic-1st}. $x(r)$ and $k$ are defined by  
\begin{align}
    x^2 &=\frac{\left(r-r_4\right)\left(r_3-r_1\right)}{\left(r-r_3\right)\left(r_4-r_1\right)}\,, \label{eq:x-definition-r} \\
    k^2 &=\frac{r_3\left(r_4-r_1\right)}{r_4\left(r_3-r_1\right)}\,.\label{eq:k-definition-r}
\end{align}
This elliptic integral expression for the angular sweep in Eq.~\eqref{eq:I_phi-integral} was originally derived in \cite{Darwin:1959-gravityfield,Chandrasekhar:1985kt}. Finally, note that the case $F(x_\rm{e};k) = 0$ means that we are at the periapsis location for the null ray. 

For practical root finding, it is useful to know a priori whether an image is direct or indirect (when root finding in Eq.~\eqref{eq:phie-integral}). Due to GR bending, the transition is no longer at $\phi_\rm{e} = 90^\circ$. The transition instead occurs when the binary is at the turning point $r_\rm{e} = r_4$, which corresponds to tangential emission
\begin{equation}
    b_\rm{max} = \frac{r_\rm{e}}{\sqrt{1 - 2M_3/r_\rm{e}}} \,.
\end{equation}
Defining $\Delta\phi_\rm{dir} \equiv \Delta\phi(r_\rm{e}, b_\rm{max})$ on the direct branch, the primary image is direct for $\phi_\rm{e} < \Delta\phi_\rm{dir}$ and indirect otherwise, while the secondary image is always indirect\footnote{For all bound b-EMRI orbits this is true because $r_\rm{e} > 4M_3$, though it is not true for plunging b-EMRI trajectories.}. A direct primary with $\phi_\rm{e} < \Delta\phi_\rm{c} \equiv \Delta\phi(r_\rm{e}, b_\rm{c})$ has $b < b_\rm{c}$. Note that the elliptic integral expressions generalize to the captured-ray case as discussed in \cite{Hackmann:2018fmk} and we use Carlson symmetric forms in scipy explicitly for this.

\subsubsection{Time delays}

To compute the time delay of the signal from a binary, we must solve the geodesic equations. In this subsection, we set $M_3 = 1$. As pointed out in Hackmann \cite{Hackmann:2018fmk}, it is possible to write this in an elliptic-integral form using Eq.~\eqref{eq:geodesic-equations-radial-roots} and the chain rule
\begin{equation}\label{eq:drdt-romer}
    \left(\frac{d r}{d t}\right)^2=\left(1-\frac{2}{r}\right)^2 \frac{b^2}{r^4} R(r) \,.
\end{equation} 
After integrating Eq.~\eqref{eq:drdt-romer}, we can invert it and express it in the following form
\begin{equation}\label{eq:delta-T-expression}
    \Delta t(r_\rm{e},b) \equiv I_t(\infty, b) \pm I_t\left(r_\rm{e}, b\right) \,,
\end{equation}
with the same sign convention as Eq.~\eqref{eq:phie-integral}, where we define
\begin{equation}\label{eq:I_t-integral}
    I_t(r,b) \equiv \int_{r_4}^{r} \frac{r^2\, d r}{b\left(1-2/r\right)\sqrt{R(r)}} \,.
\end{equation}
where we can break $I_t$ into regular and singular terms
\begin{equation}
    I_t(r,b) = I_t^\rm{reg}(r,b) + I_t^\rm{sing}(r,b) \,.
\end{equation}
The regular and singular contributions are given by 
\begin{subequations}\label{eq:Tdefs}
    \begin{align}
        I_t^\rm{reg} &=\frac{2}{\sqrt{r_4\left(r_3-r_1\right)}} \left( I_1+I_2+I_3+I_4 \right)\,,\\
        I_1 &=\left(\frac{r_3^3}{r_3-2}+\frac{1}{2}\left(r_4-r_3\right)\left(r_3-r_1+4\right)\right) F(x; k) \,,\\
        I_2 &= -\frac{1}{2} r_4\left(r_3-r_1\right) E(x; k) \, \\
        I_3 &= -2\left(r_4-r_3\right) \Pi\left(x, \frac{k^2}{c_1}; k\right) \,,\\
        I_4 &=-\frac{8\left(r_4-r_3\right)}{\left(r_4-2\right)\left(r_3-2\right)} \Pi\left(x, c_2; k\right) \,, \\
        I_t^\rm{sing} &=\frac{b \sqrt{R(r)}}{r-r_3} \\
        &\quad+ 2 \log \left(\frac{\sqrt{r\left(r-r_1\right)}+\sqrt{\left(r-r_4\right)\left(r-r_3\right)}}{\sqrt{r\left(r-r_1\right)}-\sqrt{\left(r-r_4\right)\left(r-r_3\right)}}\right) \,,
    \end{align}
\end{subequations}
where Eq.~\eqref{eq:Tdefs} is given in Ref.~\cite{Hackmann:2018fmk}, but we use the form in \cite{DellaMonica:2023ydm,DellaMonica:2025ent}. The constants $c_1,c_2$ are \cite{Hackmann:2018fmk}
\begin{subequations}
    \begin{align}
        c_1 &=\frac{r_4-r_1}{r_3-r_1} \,,\\
        c_2 &=c_1 \frac{\left(r_3-2\right)}{\left(r_4-2\right)}\,.
    \end{align}
\end{subequations}
Note that $I_t^\rm{sing}(r,b)$ is singular when $r\rightarrow\infty$, but one can regularize it via 
\begin{equation}
    I_t^\rm{sing}(r, b) = r + 2 \log r + \hat{I}_\infty^\rm{sing}(b) + \mathcal{O}\left( 1/r\right)\,,
\end{equation}
where we define
\begin{equation}\label{eq:T-infinty-hat}
    \hat I_\infty^\rm{sing}(b) \equiv r_3 + 2 \log \frac{2}{r_3+r_4} \,.
\end{equation}
Therefore, we can use the expression for $\hat I_\infty^\rm{sing}(b)$ in Eq.~\eqref{eq:T-infinty-hat} when computing $I_t^\rm{sing}(\infty, b)$ in Eq.~\eqref{eq:delta-T-expression}.

Let us now relate this to the time delay observed as the binary orbits the SMBH. The observed time delay $\Delta(t_\rm{e})$ is 
\begin{equation}\label{eq:time-delay-full}
    \Delta(t_\rm{e}) \equiv \Delta t(t_\rm{e}) - \Delta t(t_\rm{ref}) \,,
\end{equation}
where $\Delta t(t_\rm{e}) = \Delta t(r_\rm{e}(t_\rm{e}),b(t_\rm{e}))$ from Eq.~\eqref{eq:delta-T-expression} and $t_\rm{ref}$ is taken to be some reference time. It is typical to either take this to be the time at which the binary merges so that $\Delta(t_\rm{merge})=0$ or the time at which the binary crosses the ascending node (as done by Hackmann in Ref.~\cite{Hackmann:2018fmk}). Since the physical measurement of $\Delta(t_\rm{e})$ in Eq.~\eqref{eq:time-delay-full} only depends on the difference of two time delays, one can see that it is independent of the regularization procedure in Eq.~\eqref{eq:T-infinty-hat}, but we point it out since it is useful for numerical evaluation.

\begin{figure*}[t!]
    \centering
    \includegraphics[width=\textwidth]{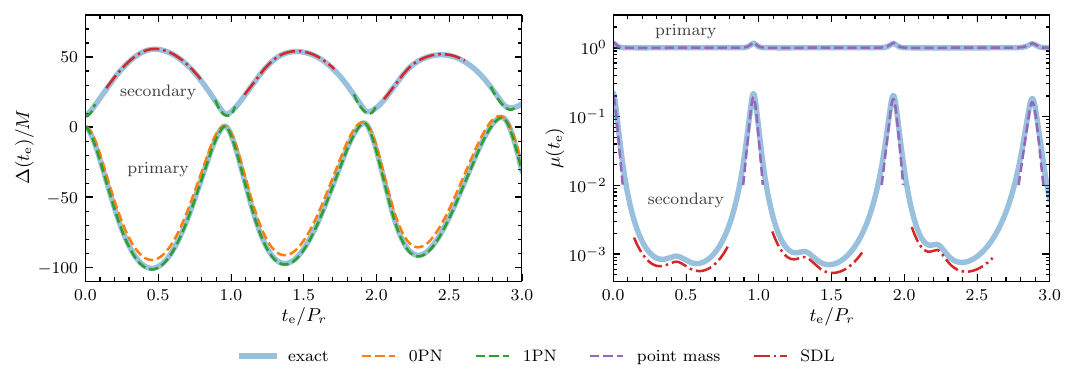}
    \caption{We compare our exact results for the time delay and magnification for a particular orbit ($a_o = 50 M_3$, $\iota_J = 72^\circ$, and $e_o = 0.3$). On the left, we show the exact time delay from Eq.~\eqref{eq:time-delay-full} (blue) versus the 0PN Rømer time delay result in Eq.~\eqref{eq:romer-time-delay-PN} (orange dashed) and 1PN result (green dashed) for primary and secondary images. For the secondary image, PN results do not exist when the source is in front of the lens, so the strong deflection limit (SDL) is used (dashed red) \cite{Bozza:2003cp}. On the right, we plot the exact magnification from raytracing (blue) computed in Eq.~\eqref{eq:magnification-close-source-app} to the point-mass limits (dashed purple) in Eq.~\eqref{eq:magnification-geometric-0PN}. We also show the secondary-image SDL magnification using the results of Ref.~\cite{Bozza:2007gt}. One can see that these time delays and magnifications agree with those computed from previous approximations with more significant deviation happening when the binary is behind the BH, where lensing is strongest.
    }
    \label{fig:time-delay}
\end{figure*}

Note that the weak-field case of the time delay $\Delta(t_\rm{e})$ is given by the Rømer time delay in the 0PN limit
\begin{equation}\label{eq:romer-time-delay-PN}
    \Delta_\rm{0PN}(t_\rm{e})=\uvec{N}_\infty \cdot \vec r_\rm{e}(t_\rm{e}) - \uvec{N}_\infty \cdot  \vec r_\rm{e}(t_\rm{ref})\,.
\end{equation}
Additional expressions are given in \cite{Hackmann:2018fmk} for the 1PN generalization of the time delay. In the left panel of Fig.~\ref{fig:time-delay}, we compare the full ray-traced time delay in Eq.~\eqref{eq:time-delay-full} to the 0PN time delay in Eq.~\eqref{eq:romer-time-delay-PN}. One can see that when the binary is behind the lens there is significant deviation between these, but this is better captured by the 1PN time delay in green. We also show that the strong deflection limit (SDL) \cite{Bozza:2003cp} for the secondary image (red) agrees with the ray-traced result.

\subsubsection{Doppler shift}

As we discussed in the main text, the Doppler shift for image $I$ is equal to 
\begin{equation}\label{eq:app-fo-fe-dtau-dte}
    \frac{f_{\rm o}^I}{f_{\rm e}} = \left.\frac{d\tau_\rm{e}}{dt_o}\right\vert_I
\end{equation}
In fact, it is more instructive to break this down into component parts
\begin{equation}
    \left.\frac{d\tau_\rm{e}}{dt_o}\right\vert_I = \left.\frac{d\tau_\rm{e}}{dt_\rm{e}}\frac{dt_\rm{e}} {dt_o}\right\vert_I\,.
\end{equation}
The Doppler shift between the binary's proper time and Schwarzschild time is given by
\begin{equation}
    \left.\frac{d\tau_\rm{e}}{dt_\rm{e}}\right\vert_I = \frac{1-2M_3/r_\rm{e}}{\tilde E} 
\end{equation}
where $\tilde E$ is the conserved energy of the bound orbit as discussed in App.~\ref{app:motion-precession}. It is more instructive to write this in a different form where we imagine a static observer at the COM of the binary which measures the binary passing with speed $v$. In this case, we can express the time delay as
\begin{equation}\label{eq:dtau-dte-gamma-gravitational}
    \left.\frac{d\tau_\rm{e}}{dt_\rm{e}}\right\vert_I = \frac{1}{\gamma(v)} \sqrt{1 - \frac{2 M_3}{r_\rm{e}}}
\end{equation}
where $v$ is the velocity of the binary's COM in the static frame and $\gamma(v) = 1/ \sqrt{1-v^2}$ is the Lorentz factor. We have only repackaged the results for Schwarzschild because the conservation law relates position and velocity $\tilde E = \gamma(v)\sqrt{1-2 M_3/r_\rm{e}}$. Eq.~\eqref{eq:dtau-dte-gamma-gravitational} explicitly packages the transverse redshift $\gamma(v)$ and the gravitational redshift $\sqrt{1 - \frac{2 M_3}{r_\rm{e}}}$ separately. We can play the same game for the other component of the chain rule and show 
\begin{equation}\label{eq:dte-dto-romer}
    \left.\frac{dt_\rm{e}}{dt_o}\right\vert_I = \frac{1}{1 - v_{\parallel}^I}
\end{equation}
where $v_\parallel^I \equiv \uvec k_0^I \cdot \vec v$ is the component of $\vec v$ which is parallel to $\uvec k_0^I$. One can show that this definition is equivalent to $v_\parallel^I = - \Delta'_I(t_\rm{e})$. Therefore, Eq.~\eqref{eq:dte-dto-romer} encodes the longitudinal redshift analogous to the Rømer time delay. Thus, the ratio of observed frequency to source frame frequency is equal to 
\begin{equation}
    \frac{f_{\rm o}^I}{f_{\rm e}} = \sqrt{1 - \frac{2 M_3}{r_\rm{e}}} \frac{1}{\gamma(v)} \frac{1}{1 - v_\parallel^I}
\end{equation}
where the first term is the gravitational redshift, the second is the transverse Doppler shift, and the third is the longitudinal Doppler shift.

\subsubsection{Magnification}

In this work, we use the lensed magnification for a source arbitrarily close to the Schwarzschild BH. The magnification of a binary which is sitting in the static tetrad was originally computed in \cite{Frittelli:1999yf,Beloborodov:2002mr,Poutanen:2019tcd}. From Eq.~(14)\footnote{One needs to identify parameters in this paper as $\mathcal{D} \rightarrow \mu$, $\psi \rightarrow \phi_\rm{e}$, $\alpha \rightarrow \beta_\rm{e}$, and $u \rightarrow 2 M_3 /r_\rm{e}$ of \cite{Poutanen:2019tcd} to get Eq.~\eqref{eq:magnification-poutanen}.} of \cite{Poutanen:2019tcd}, the magnification is given by 
\begin{equation}\label{eq:magnification-poutanen}
    \mu(r_\rm{e},b) = \left( 1 - \frac{2 M_3}{r_\rm{e}} \right)^{-1} \frac{d \cos \beta_\rm{e}}{d\cos\phi_\rm{e}}\,,
\end{equation}
where $\cos \beta_\rm{e} \equiv \uvec k_0 \cdot \uvec r_\rm{e}$. $\beta_\rm{e}$ expresses how tangentially the geodesic is emitted and is given explicitly later in Eq.~\eqref{eq:sin-beta-equation}. We can convert this into a more useful form
\begin{equation}\label{eq:magnification-close-source-app}
    \mu(r_\rm{e},b) = \frac{1}{\sqrt{R(r_\rm{e};b)} \left|\sin \phi_\rm{e}\right|\left|\partial_b \phi_\rm{e}\right|}\,,
\end{equation}
where we used $|\cos\beta_\rm{e}| = b\sqrt{R(r_\rm{e};b)}/r_\rm{e}^2$ and worked through the algebra. This form is extremely useful because it works for arbitrarily close source locations. Note that we can also make the magnification more interpretable by expressing it as
\begin{equation}\label{eq:magnification-projective-illustrative}
    \mu(r_\rm{e},b) = \frac{b}{r_\rm{e}^2 |\cos \beta_\rm{e}| \left|\sin \phi_\rm{e}\right|\left|\partial_b \phi_\rm{e}\right|}\,.
\end{equation}
This reduces to the strong-deflection thin-lens expressions in \cite{Ohanian:1987xga} (i.e.~when $|\cos \beta_\rm{e}| \rightarrow 1$ and $r_\rm{e} \gg M_3$). 

Our expressions reduce to the standard weak-field, thin-lens point-mass magnification used in GW lensing \cite{Schneider:1992bmb,Nakamura:1999uwi,Takahashi:2003ix} when $b \gg M_3$, $r_\rm{e} \gg M_3$ and the source is behind the lens ($\cos \phi_\rm{e}<0$)
\begin{equation}\label{eq:magnification-geometric-0PN}
    \mu_{1,2}^\rm{pm} = 1/2 \pm \frac{\eta^2+2}{2\eta(\eta^2+4)^{1/2}} \,,
\end{equation}
with $\eta = r_\rm{e}|\sin\phi_\rm{e}|/(2\sqrt{- M_3 r_\rm{e}\cos\phi_\rm{e}})$ and where the negative sign of $\mu_2^\rm{pm}$ reflects the parity of the secondary image. The time delay between images is given by 
\begin{equation}\label{eq:image-time-delay-0PN}
    \Delta t_l = 4M_3\left[\frac{\eta\sqrt{\eta^2+4}}{2} + \ln\frac{\sqrt{\eta^2+4}+\eta}{\sqrt{\eta^2+4}-\eta}\right] \,.
\end{equation}
Note that our strong-field magnification formula applies to every image, with the impact parameter of each obtained from the lens equation. In the right panel of Fig.~\ref{fig:time-delay}, we show that the strong-field magnification expressions agree with the point-mass results Eq.~\eqref{eq:magnification-geometric-0PN}. We also show the secondary image magnification agrees with the ones developed for SDL~\cite{Bozza:2007gt}.

The GW signal is magnified due to (a) lensing and (b) Doppler shifts. As described in \cite{Cusin:2024git}, the Doppler shift magnification is given by the ratio of the observed frequency to emitted frequency. In Eq.~\eqref{eq:app-fo-fe-dtau-dte}, we provide a convenient form for understanding this magnification from the Doppler shift, and it is understood that this is applied on top of Eq.~\eqref{eq:magnification-close-source-app} expression for the lensing magnification.

\subsection{Matching to binary's source frame}\label{app:matching-source-frame}

In this section, we explain how the GW that reaches an observer at infinity is modified due to how LOS modulations change the matching to the binary's source frame. This is done in the following order: (1) deriving how precession of $\uvec L_i$ modulates the waveform in flat space, (2) describing how lensing modulates the LOS, (3) explaining aberration LOS modulation in flat space, and (4) combining all three self-consistently. For simplicity, we will drop all image $I$ labels so that the notation is cleaner until Sec.~\ref{app:combining-lensing-aberration}.

\subsubsection{Review of precession modulations in flat space}\label{app:review-precession}

In this manuscript, we compute how precession modulates the waveform and it is useful to derive this in the flat space case as was done in ACST \cite{Apostolatos:1994mx}. Generically we have an orbital separation vector $\uvec r$ in which phase is measured from `principal+' direction $\uvec p_\rm{flat}$
\begin{equation}
    \uvec r = \cos \Phi \,\uvec p_\rm{flat} + \sin\Phi\, \uvec q_\rm{flat}
\end{equation}
where 
\begin{equation}
    \uvec p_\rm{flat} \equiv \frac{\uvec L_i \times \uvec N_\infty}{\Vert \uvec L_i \times \uvec N_\infty \Vert} \,,
\end{equation}
and $\uvec q_\rm{flat} \equiv \uvec L_i \times \uvec p_\rm{flat}$ is the other one. In the rest of this SM, we will allow $\uvec N$ to be varying due to the effects of lensing bending and aberration, but we will consider this fixed here in the flat space limit\footnote{Analogously, we match the notation of \cite{Apostolatos:1994mx} and \cite{Yu:2020dlm} where $\uvec N = \uvec N_\rm{src} = \uvec N_\infty$, which is valid when the binary is far from the SMBH.}. Due to parallel transport, the inner angular momentum is time dependent and precesses in a cone of opening angle $\lambda_L$ about $\uvec J$ \cite{Akcay:2016dku,Yu:2020dlm}. This evolves as 
\begin{equation}
    \frac{d \uvec L_i}{dt} = \vec \Omega_\rm{prec} \times \uvec L_i \,,
\end{equation}
where the precession frequency is given by $\vec \Omega_\rm{prec} \equiv \dot \alpha \uvec J$. This solution can be explicitly written in coordinates via 
\begin{equation}\label{eq:desitter-precession}
    \uvec L_i(t) = \cos \lambda_L \uvec J + \sin \lambda_L \left[ \cos \alpha(t) \uvec n_\Omega + \sin \alpha(t)\uvec m \right]\,,
\end{equation}
where $\uvec n_\Omega, \uvec m$ are two node vectors\footnote{These can be arbitrarily chosen, but one useful construction is $\uvec n_\Omega$ as the intersection of the ecliptic plane and the orbital plane.} chosen perpendicular to $\uvec J$ and $\alpha$ is the precession phase measured in the orbital plane from $\uvec n_\Omega$ with $\alpha(0)=\alpha_0$. 

One can see that time-varying $\uvec L_i$ in Eq.~\eqref{eq:desitter-precession} will change the reference axis $\uvec p_\rm{flat}$ that our waveform is defined against, which will give rise to the Thomas phase modulation. This can be derived by taking the time derivative of $\uvec p_\rm{flat} \cdot \uvec r = \cos \Phi$, which yields 
\begin{equation}
    \delta \dot \Phi = -\frac{\dot{\uvec p}_\rm{flat} \cdot \uvec r}{\sin \Phi} + \dot \alpha \cos \lambda_L
\end{equation}
where in this derivation we used that the derivative of $\uvec r$ with respect to $t$ is given by
\begin{align}\label{eq:rdot-acst-comparison}
    \frac{d \uvec r}{dt} &= \left( \dot{\Phi}_\rm{int} \uvec L_i + \vec \Omega_\rm{prec} \right) \times \uvec r \,,
\end{align}
where the first term represents the rate at which the binary rotates in the local frame (cf.~$\vec \Omega_\rm{orb} \equiv \dot{\Phi}_\rm{int} \uvec L_i$) and the second term is the rate of the external frame rotation due to de Sitter precession/Fermi--Walker transport. Explicitly taking the time derivative of $\uvec p$ and utilizing Eq.~\eqref{eq:desitter-precession} yields 
\begin{equation}
    \delta \dot \Phi = \left( \frac{\uvec L_i \cdot \uvec N_\infty}{1 - \left( \uvec L_i \cdot \uvec N_\infty \right)^2} \right)\left( \uvec L_i \times \uvec N_\infty \right) \cdot\dot{\uvec L_i} + \dot \alpha \uvec L_i \cdot \uvec J \,,
\end{equation}
which matches ACST Eq.~(28) up to a neglected factor of $\dot \alpha \uvec L_i \cdot \uvec J  =\dot \alpha \cos\lambda_L$ which is the coprecessing phase. Therefore the Thomas phase is given by the integral
\begin{equation}\label{eq:thomas-phase-integral}
    \Phi_T(t) = - \int_{t}^{t_c} \frac{\bm{\Omega}_\rm{prec} \cdot \left[ \uvec L_i - (\uvec L_i \cdot \uvec N_\infty)\uvec N_\infty \right]}{1-\left( \uvec L_i \cdot \uvec N_\infty \right)^2}\, dt \,,
\end{equation}
where we normalize $\Phi_T(t_c) = 0$ by absorbing it into the phase of coalescence $\phi_c$ as is done in \cite{Apostolatos:1994mx}. We note that the additional coprecessing phase was neglected in some previous hierarchical triple works \cite{Yu:2020dlm,Yu:2021dqx,Laeuger:2023qyz} which we correct here. We note that this in-plane precession contribution is not new and is known in the waveform modeling community \cite{Arun:2008kb,Schmidt:2010it,Boyle:2011gg,Schmidt:2012rh} and was neglected in ACST \cite{Apostolatos:1994mx} due to its focus on tilt precession over in-plane precession. This additional term does not change the qualitative features of how much precession modulates the waveform, but just redistributes which lines of sight have the greatest modulation. The Thomas phase has both an oscillatory and a secular contribution, the latter of which corresponds to the number of windings. After a complete precession timescale, the accumulated Thomas phase is equal to
\begin{equation}\label{eq:thomas-winding-number}
    \Delta \Phi_T^{\rm{ cycle}}= \begin{cases}
        2 \pi \operatorname{sgn}\left(\cos \lambda_L\right), & \left|\cos \lambda_L\right|>|\uvec J \cdot \uvec N_\infty | \,, \\ 
        0, & \left|\cos \lambda_L\right|<|\uvec J \cdot\uvec N_\infty| \,.
    \end{cases}
\end{equation}
Equation~\eqref{eq:thomas-winding-number} represents the number of windings when the viewing direction is outside the cone that $\uvec L_i$ traces. When viewed inside the cone $\left|\cos \lambda_L\right|<|\uvec J \cdot\uvec N_\infty|$, there is only an oscillatory Thomas phase and no secular term.

Note that the Thomas phase can be understood geometrically with spherical trigonometry in contrast to the integral expression in Eq.~\eqref{eq:thomas-phase-integral}. In the corotating/coprecessing frame $(\uvec e_1^\rm{cp}, \uvec e_2^\rm{cp}, \uvec L_i)$\footnote{$\uvec e_1^\rm{cp}$ can be generically chosen, as long as it rotates with $\dot{\uvec e}_1^\rm{cp} = \vec \Omega_\rm{prec} \times \uvec e_1^\rm{cp}$ and $\uvec e_2^\rm{cp} = \uvec L_i \times \uvec e_1^\rm{cp}$. Here we pick $\uvec e_1^\rm{cp} \propto \uvec J \times \uvec L_i$ for convenience.} \cite{Buonanno:2002fy,Boyle:2011gg}, the binary evolves as if it were an isolated binary
\begin{equation}
    \uvec r = \cos \Phi_\rm{int} \uvec e_1^\rm{cp} + \sin \Phi_\rm{int}\uvec e_2^\rm{cp} \,.
\end{equation}
The Thomas phase is thus the amount of additional rotation that occurs when we measure the phase with respect to the principal+ axis $\uvec p$ compared to this inertial frame. The observed phase accumulation is given by 
\begin{equation}
    \Phi_\rm{obs}(t) = \Phi_\rm{int}(t) + \Phi_T^\rm{prec}(t)\,,
\end{equation}
where the Thomas phase $\Phi_T^\rm{prec}$ is the angle between $\uvec p$ and $\uvec e_1^\rm{cp}$ 
\begin{align}\label{eq:thomas-phase-spherical-trig}
    \Phi_T^\rm{prec}(t) &= \angle_{\uvec L_i}\left(\uvec p_\rm{flat} \rightarrow \uvec e_1^\rm{cp} \right) \nn\\
    &= \operatorname{atan2}\left[ \uvec q_\rm{flat} \cdot \uvec e_1^\rm{cp}, \uvec p_\rm{flat} \cdot \uvec e_1^\rm{cp} \right] + 2 \pi n\,,
\end{align}
where the branch $n$ is chosen so that $\Phi_T^\rm{prec}(t)$ is continuous, we set $\Phi_T^\rm{prec}(t_c) = 0$, and $\angle_{\uvec a}(\uvec u, \uvec v) \equiv \operatorname{atan2} \left[ \uvec a \cdot \left( \uvec u \times\uvec v \right), \uvec u \cdot \uvec v \right]$ is the counterclockwise angle from $\uvec u$ to $\uvec v$ around $\uvec a$. Note that Eq.~\eqref{eq:thomas-phase-spherical-trig} is algebraic as formulated, so it can be computed quickly without implementing a history integral in the case of simple precession where $\vert \vec L_i \vert \ll \vert \vec J\vert$. We use Eq.~\eqref{eq:thomas-phase-spherical-trig} in our code for computing precession because the problem is more simply formulated, though we use the explicit definition in Eq.~\eqref{eq:thomas-phase-prec-atan2-useful}.

Additionally, the precession modulates the inclination and polarization phase of the observed gravitational wave at infinity. The inclination is rather trivially just modified to be time-dependent via 
\begin{equation}
    \cos \iota_\rm{flat}(t) = \uvec L_i \cdot \uvec N_\infty \,.
\end{equation}
The polarization phase is modulated in a more complicated manner. This can be understood most easily by switching to the J-frame \cite{Pratten:2020ceb} where our polarization angle is measured in reference to a fixed direction $\uvec e_{\uvec J}$ which depends on the total angular momentum 
\begin{equation}
    \uvec e_{\uvec J } = \left[ \uvec J - \left( \uvec J \cdot \uvec N_\infty \right) \uvec N_\infty \right]/\Vert ... \Vert \,,
\end{equation}
where it is normalized accordingly. This J-frame is typically used in waveform modeling, and reads the polarization off from the vector $\uvec e_{\uvec J}$ which points towards the north pole of $\uvec J$. In this frame, we can show that the polarization direction is expressed as
\begin{equation}\label{eq:Jframe-prec}
    \uvec p_\rm{flat} = \cos \psi^\rm{prec} \uvec e_{\uvec J} + \sin \psi^\rm{prec} \left( \uvec N_\infty \times \uvec e_{\uvec J} \right)\,.
\end{equation}
Eq.~\eqref{eq:Jframe-prec} allows us to express the rotation of the polarization angle $\psi^\rm{prec}$ as
\begin{equation}
    \psi^\rm{prec} =\angle_{\uvec N_\infty}\left( \uvec e_{\uvec J} \rightarrow \uvec p_\rm{flat} \right)\,.
\end{equation}
We stress that the expression for the polarization phase rotation $\psi^\rm{prec}$ is computed for \textit{an observer at infinity} perpendicular to $\uvec N_\infty$ while the Thomas phase rotation $\Phi_T^\rm{prec}$ is computed as a rotation in the \textit{LOS in the binary's source frame} around $\uvec L_i$. 

By using the J-frame, we express the precessional polarization rotation separately from the polarization rotation due to space-based detectors. Note that in some previous works on LISA and triple waveforms, the polarization phase is defined via the detector's polarization axis \cite{Cutler:1997ta,Yu:2020dlm}. This is important because the LISA detector's normal direction rotates, so the observed polarization in LISA is given by 
\begin{equation}
    \psi^\rm{det} = \Delta \psi^\rm{det} + \psi_J^{\rm{prec}} \,,
\end{equation}
where all angles are measured counterclockwise about $\uvec N$, and
\begin{equation}
    \Delta\psi^\rm{det}(t) = \operatorname{atan2}\left[\uvec N_\infty\cdot\left(\uvec e_1^\rm{det}(t)\times\uvec e_{\uvec J}\right), \uvec e_1^\rm{det}(t)\cdot\uvec e_{\uvec J}\right]\,,
\end{equation}
where we define $\uvec e_1^\rm{det} = \left[ \uvec z^\rm{det} - \left( \uvec N_\infty \cdot \uvec z^\rm{det} \right)\uvec N_\infty \right] / \Vert ... \Vert $ and $\uvec e_2^\rm{det} = \uvec N_\infty \times \uvec e_1^\rm{det}$ which agrees with \cite{Cutler:1997ta,Yu:2020dlm} conventions for polarization phase. Note that all source frame time-dependent quantities depend on $\tau_\rm{e}(t_\rm{e}^I(t_o))$ while a rotating detector like LISA has a normal detector orientation given by $\uvec z^\rm{det}(t_o)$. All polarization phases are only defined up to a multiple of $\pi$ in this Supplemental Material. In this Letter, we do computations in the J-frame to compare with Teukolsky b-EMRI results in the main text \cite{Santos:2025ass}, but the detector frame is important for use with real space-based detector data.

\subsubsection{Lensing modulations}

To compute how bending modifies the waveform, we follow the construction in \cite{Beloborodov:2002mr,Poutanen:2019tcd}. As shown in \cite{Beloborodov:2002mr}, the angle between the radial direction and the null geodesic trajectory in the static frame is given by 
\begin{equation}\label{eq:sin-beta-equation}
    \sin \beta_\rm{e} = \frac{b}{r_\rm{e}} \sqrt{1 - \frac{2 M_3 }{r_\rm{e}}}
\end{equation}
where $\cos \beta_\rm{e} = -\uvec N_0\cdot \uvec r_\rm{e} $ is defined analogously for a null geodesic traveling in direction $-\uvec N_0$ in the static frame\footnote{$\beta_\rm{e}$ is computed from this formula but with the sign unwrapped depending on whether it is inwardly or outwardly propagating.}. The results of \cite{Poutanen:2019tcd} give a convenient expression for the local null geodesic direction as 
\begin{equation}\label{eq:N0-equation-poutanen}
    \uvec N_0 = \frac{1}{\sin \phi_\rm{e}} \left[ \sin \beta_\rm{e} \uvec N_\infty - \sin \left( \delta \right)\uvec r_\rm{e} \right]\,,
\end{equation}
where $\delta \equiv \phi_\rm{e}  - \beta_\rm{e}$ is the amount of bending that the geodesic has experienced. 

Since Schwarzschild geodesics are planar (i.e.~have no twists), Eq.~\eqref{eq:N0-equation-poutanen} can be expressed as a rotation matrix formula 
\begin{align}
    \uvec N_0 &= \mat R_\rm{lens} \uvec N_\infty \,, \label{eq:N0-equation-rotation}\\
    \mat R_\rm{lens} &\equiv \mat R_{\uvec i}(-\delta) \,,\nn
\end{align}
where $\uvec i$ is the direction of the null ray's angular momentum which is conserved and perpendicular to the $(\uvec r_\rm{e}, \uvec N_\infty)$ plane,
\begin{equation}
    \uvec i \equiv \pm\frac{\uvec N_\infty \times \uvec r_\rm{e}}{\Vert \uvec N_\infty \times \uvec r_\rm{e}\Vert}\,.
\end{equation}
with $+$ for primary image and $-$ for secondary image.
Therefore, we have constructed a unit basis near the null ray $(\uvec i, \uvec j(\lambda),\uvec N(\lambda))$ which has simple solutions and is parallel transported. The other direction $\uvec j$ is parallel transported via the same rotation as Eq.~\eqref{eq:N0-equation-rotation}, i.e., 
\begin{equation}
    \uvec j_0 \equiv \uvec i \times \uvec N_0 = \mat R_\rm{lens} \uvec j_\infty\,.
\end{equation}

Therefore, a generic polarization direction at infinity can be related to the polarization direction at the binary in the static tetrad as 
\begin{equation}
    \uvec p_\infty^\rm{lens} = \mat R_\rm{lens}^T \uvec p_0^\rm{lens}\,,
\end{equation}
where $\uvec p_0^\rm{lens} \equiv \uvec L_i \times \uvec N_0 / \Vert ... \Vert$ is the polarization axis at emission taking into account bending, and $\uvec p_\infty^\rm{lens}$ is the polarization axis at infinity that is rotated due to propagation. Therefore the effect of lensing on the LOS and principal+ polarization axis is simply expressed through rotation matrices, which gives us a framework to match the GW from the binary to the one observed at infinity.

\subsubsection{Aberration modulations}

In addition to the effect of lensing, there is relativistic aberration that occurs due to the binary moving with velocity $\vec v$ with respect to the static tetrad. Luckily, since we are located at the same place in the manifold, relativistic aberration is given by the framework developed in special relativity. In this section, we primarily follow the foundational works of \cite{Bonvin:2022mkw,He:2022qhc,Cusin:2024git}. Relativistic aberration looks like 
\begin{equation} \label{eq:usual-aberrated-formula}
    \uvec N_\rm{src} = \frac{1}{\mathcal{D}} \left\{ \uvec N_0 + \left[ \left( \gamma - 1 \right) \uvec N_0 \cdot \uvec v + \gamma v \right]\uvec v \right\}\,,
\end{equation}
where $\vec v$ is the center-of-mass velocity of the binary in the static frame, $\uvec v = \vec v / v$, $\uvec N_\rm{src}$ is the aberrated direction in the binary's frame, and $\gamma = 1/\sqrt{1-v^2}$ is the Lorentz factor. The Doppler factor is defined to be
\begin{equation}
    \mathcal{D} = \gamma \left( 1 + \uvec N_0 \cdot \vec v \right) \,.
\end{equation}
As mentioned in App.~B of \cite{Cusin:2024git}, this operation is best understood as a rotation matrix, which we define as 
\begin{equation}
    \mat R_\rm{ab} \equiv \mat R_{\uvec a}(\epsilon)\,,
\end{equation}
where the aberration angle is $\epsilon = \tilde \theta -  \theta$ and $\cos \theta = -\uvec N_0 \cdot \uvec v$ and $\cos \tilde \theta = -\uvec{ N}_\rm{src} \cdot \uvec{v}$ are the directions the velocity is moving along. The rotation axis $\uvec a$ is given by 
\begin{equation}
    \uvec a = \frac{\uvec N_0 \times \vec v}{|\uvec N_0 \times \vec v|} \,.
\end{equation}
The results of Eq.~\eqref{eq:usual-aberrated-formula} can thus be compactly rephrased as 
\begin{equation}
    \uvec N_\rm{src} = \mat R_\rm{ab} \uvec N_0\,,
\end{equation}
where the amount of aberration is given by $\cos \epsilon = \uvec N_\rm{src} \cdot \uvec N_0$. 
As discussed in \cite{Cusin:2024git}, the same rotation matrix carries the polarization basis between the two frames. Therefore, we can relate the polarization direction in the binary's frame $\uvec p_\rm{src}$ to the one in the static tetrad $\uvec p_0 $ via 
\begin{equation}\label{eq:p0-static-tetrad}
    \uvec p_0 = \mat R_\rm{ab}^T \uvec p_\rm{src}  \,.
\end{equation}
where $\uvec p_\rm{src} \equiv \uvec L_i \times \uvec N_\rm{src}/\Vert ... \Vert$. This will allow us to combine the effects of lensing plus aberration when constructing our triple waveform at infinity.

\subsubsection{Combining the effects of lensing and aberration}\label{app:combining-lensing-aberration}

Let us now explain how to stack the effects of lensing and aberration for GW emitted in a triple. This has been studied for the cases of EM emission in some previous works \cite{Poutanen:2020lom,Loktev:2021nhk} which explain how the EM polarization is rotated, but we also carefully track how it matches the source frame emission and how the Thomas phase is modified in GW. As we have shown in the previous two sections, the effects of lensing and aberration modify the waveform through a combination of rotation matrices. In the rest of this section, we add the image index $I$ again to distinguish between each ray. Both aberration and lensing stack and can compactly be written as
\begin{align}
    \uvec N_\rm{src}^I &= \mat R_{\rm{ab},I} \mat R_{\rm{lens},I} \uvec N_\infty \,, 
\end{align}
where $\uvec N_\rm{src}^I = - \uvec k_\rm{src}^I$ is the LOS direction in the binary's frame. At emission, the $\uvec p_\rm{src}^I$ principal+ polarization axis is defined in the source frame with the usual definition $\uvec p_\rm{src}^I \equiv  \uvec L_i \times \uvec N_\rm{src}^I /\Vert ...\Vert$. As we showed in the previous two sections, the principal+ polarization axis at $\infty$ is given by multiplying the lensing rotation matrix times the static tetrad's polarization $\uvec p_0$ from Eq.~\eqref{eq:p0-static-tetrad}
\begin{equation}\label{eq:polarization-rotation-total}
    \uvec p_\infty^I = \mat R_{\rm{lens},I}^T \mat R_{\rm{ab},I}^T \uvec p_\rm{src}^I\,.
\end{equation}
Before we do this, we will define some useful expressions for polarization directions at emission and at infinity
\begin{gather}\label{eq:p-em-infty}
    \uvec p_\rm{flat} \equiv \frac{\uvec L_i \times \uvec N_\infty}{\Vert \uvec L_i \times \uvec N_\infty \Vert } \,, \\
    \begin{aligned}
        \uvec p_0^{\rm{lens},I} &\equiv \frac{\uvec L_i \times \uvec N_0^I}{\Vert \uvec L_i \times \uvec N_0^I \Vert} \,,
        & \uvec p_\infty^{\rm{lens},I} &\equiv \frac{\uvec L_\rm{lens}^I \times \uvec N_\infty}{\Vert \uvec L_\rm{lens}^I \times \uvec N_\infty \Vert} \,, \\
        \uvec p_\rm{src}^I &\equiv \frac{\uvec L_i \times \uvec N_\rm{src}^I}{\Vert \uvec L_i \times \uvec N_\rm{src}^I \Vert } \,,
        & \uvec p_\infty^I &\equiv \frac{\uvec L_\rm{eff}^I \times \uvec N_\infty}{\Vert \uvec L_\rm{eff}^I \times \uvec N_\infty \Vert} \,,
    \end{aligned} \nn
\end{gather}
where $\uvec L_\rm{lens}^I = \mat R_{\rm{lens},I}^T \uvec L_i$ and $\uvec L_\rm{eff}^I = \mat R_{\rm{lens},I}^T \mat R_{\rm{ab},I}^T \uvec L_i$. For flat space, the polarization direction at emission and at infinity are the same\footnote{i.e.~$\uvec p_0^\rm{flat} = \uvec p_\infty^\rm{flat} = \uvec p_\rm{flat}$}, while the `lens' line includes contributions from bending of the null ray during propagation, and the final line includes aberration and bending together. Note that both $\uvec p_\rm{flat}, \uvec p_0^{\rm{lens},I},\uvec p_\rm{src}^I \perp \uvec L_i$ and  $\uvec p_\rm{flat}, \uvec p_\infty^{\rm{lens},I},\uvec p_\infty^I \perp \uvec N_\infty$ by construction, so angles between them simply add. Grouping the polarization directions as in Eq.~\eqref{eq:p-em-infty} allows us to generalize how the inclination angle, Thomas phase, and polarization angle are modulated from these effects.

Just as we saw in the case where $\uvec N = \uvec N_\infty$ is fixed, with no rotation due to ignoring lensing and aberration, the generalization of the phasing formula is given by 
\begin{equation}
    \Phi_{\rm{obs}}^I(t_o^I(\tau)) = \Phi_\rm{int}(\tau) + \Phi_{T,I}(\tau) + \frac{\pi n_I}{4}\,,
\end{equation}
where the total Thomas phase $\Phi_{T,I}$ is given by the angle between the principal+ axis $\uvec p_\rm{src}^I $ and the parallel-transported coprecessing basis $\uvec e_1^\rm{cp}$ and $n_I=0,1$ is the Morse index for a type I and type II image, respectively \cite{Takahashi:2003ix,Dai:2017huk,Ezquiaga:2020gdt}. The expression for the Thomas phase of image $I$ is given by 
\begin{equation}\label{eq:app-thomas-phase-breakup}
    \Phi_{T,I} \equiv \angle_{\uvec L_i}\left(\uvec p_\rm{src} \rightarrow\uvec e_1^\rm{cp} \right) = \Phi_{T,I}^\rm{prec} + \Delta\Phi_{T,I}^\rm{lens} + \Delta\Phi_{T,I}^\rm{ab}\,.
\end{equation}
These individual terms in Eq.~\eqref{eq:app-thomas-phase-breakup} are thus defined as rotations in the orbital plane $\perp \uvec L_i$ using emission polarization direction definitions in Eq.~\eqref{eq:p-em-infty} 
\begin{subequations}
    \begin{align}
        \Phi_{T,I}^\rm{prec} &\equiv \angle_{\uvec L_i}\left(\uvec p_\rm{flat} \rightarrow \uvec e_1^\rm{cp} \right) \,,\\
        \Delta\Phi_{T,I}^\rm{lens} &\equiv \angle_{\uvec L_i}\left( \uvec p_0^\rm{lens} \rightarrow \uvec p_\rm{flat}\right) \,,\\
        \Delta\Phi_{T,I}^\rm{ab} &\equiv \angle_{\uvec L_i}\left( \uvec p_\rm{src} \rightarrow \uvec p_0^\rm{lens} \right) \,.
    \end{align}
\end{subequations}
These can be explicitly written with the unwrapped arctan 
\begin{widetext}
    \begin{subequations}\label{eq:thomas-phase-atan2-all}
        \begin{align}
            \Phi_{T,I}^\rm{prec} &= \operatorname{atan2}\left[ 
                -\uvec N_\infty \cdot \left( \uvec J \times \uvec L_i \right), 
                \left( \uvec J \cdot \uvec L_i \right)\left( \uvec L_i \cdot \uvec N_\infty \right) - \uvec J \cdot \uvec N_\infty 
                \right] + 2 \pi n\,, \label{eq:thomas-phase-prec-atan2-useful}\\
            \Delta\Phi_{T,I}^\rm{lens} &=\operatorname{atan2} \left[ 
                -\uvec L_i \cdot \left( \uvec N_\infty \times \uvec  N_0 \right) , 
                \uvec N_\infty \cdot \uvec N_0 - \left( \uvec L_i \cdot \uvec N_\infty \right)\left( \uvec L_i \cdot \uvec N_0 \right) 
                \right] \,,\\
            \Delta\Phi_{T,I}^\rm{ab} &=\operatorname{atan2} \left[ 
                -\uvec L_i \cdot \left( \uvec N_0 \times \uvec  N_\rm{src} \right) , 
                \uvec N_0 \cdot \uvec N_\rm{src} - \left( \uvec L_i \cdot \uvec N_0 \right)\left( \uvec L_i \cdot \uvec N_\rm{src} \right) 
                \right] \,,
        \end{align}
    \end{subequations}
\end{widetext}
where each of these is evaluated at the parameters associated with image $I$. Note that the first term for $\Phi_{T,I}^\rm{prec}$ matches the flat space result in Eq.~\eqref{eq:thomas-phase-spherical-trig}; however, it still depends on image $I$ through the implicit time-delay dependence of $\uvec L_i(t_o^I(\tau))$. The measured inclination is given by 
\begin{equation}\label{eq:cos-iota-I}
    \cos \iota_I = \uvec N_\infty \cdot \uvec L_\rm{eff}^I \,.
\end{equation}
Therefore, the LOS modulations from how the ray arrives at the source plane are characterized by inclination changes in Eq.~\eqref{eq:cos-iota-I} and modulations to phase are in Eqs.~\eqref{eq:thomas-phase-atan2-all}. 

\begin{figure*}[t]
    \centering
    \includegraphics{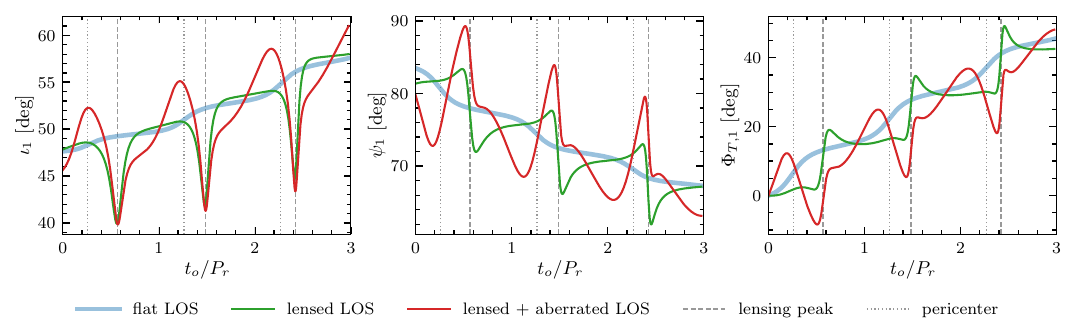}
    \caption{We show modulations to the inclination, polarization angle, and Thomas phase for the primary image in a triple. In each plot, we show how the flat LOS (blue), lensed LOS (green), and lensed + aberrated LOS change each of these parameters. We use a waveform with $a_o = 50 M_3$, $e_o=0.3$, $\iota_J = 72^\circ$, and $\lambda_L = 25^\circ$. This is plotted for $t_\rm{o}$ up to 3 radial orbits in $P_r$, which corresponds to about $10\%$ of a precession timescale ($3 P_r \approx 0.1 P_\rm{prec}$). We plot where the outer orbit has its pericenter (dotted grey) and lensing peak (dashed grey). One can see for the flat LOS case, there is angular momentum precession occurring, and its rate is fastest near pericenter. Additionally, one can see that throughout an outer orbit, the LOS modulation changes very rapidly near peak lensing and near pericenter because alignment and aberration are larger at these points.}
    \label{fig:iota-psi-phiT-primary}
\end{figure*}

In addition to LOS modulations, the polarization angle is changed in our triple system due to precession, lensing and aberration. An observer would see the polarization rotated according to 
\begin{align}
    \psi_I &\equiv \angle_{\uvec N_\infty}\left( \uvec e_{\uvec J} \rightarrow \uvec p_\infty^I \right)\,,\nn \\
    &= \psi^\rm{prec} + \Delta \psi^\rm{lens}_I + \Delta \psi^\rm{ab}_I \,, \label{eq:polarization-sum-app}
\end{align}
where each term in Eq.~\eqref{eq:polarization-sum-app} is constructed by defining some `fictitious' polarization directions at infinity using definitions in Eq.~\eqref{eq:p-em-infty}. The different polarization angle rotations are defined as angles measured around $\uvec N_\infty$ at infinity
\begin{align}\label{eq:pol-phase-angles}
    \psi^\rm{prec} &\equiv \angle_{\uvec N_\infty}\left( \uvec e_{\uvec J} \rightarrow \uvec p_\infty^\rm{flat} \right)\,, \nn\\
    \Delta \psi^\rm{lens}_I &\equiv \angle_{\uvec N_\infty}\left( \uvec p_\infty^\rm{flat} \rightarrow \uvec p_\infty^{\rm{lens},I} \right) \,,\\
    \Delta \psi^\rm{ab}_I &\equiv \angle_{\uvec N_\infty}\left( \uvec p_\infty^{\rm{lens},I} \rightarrow \uvec p_\infty^I \right)\,,\nn
\end{align}
where each of these is the angle measured around $\uvec N_\infty$. The expressions in Eq.~\eqref{eq:pol-phase-angles} are additive since the `fictitious' polarization directions at infinity are all perpendicular to $\uvec N_\infty$. These can be explicitly evaluated as
\begin{subequations}\label{eq:full-polarization-eqns}
    \begin{align}
        \tan\psi^{\rm{prec}} &= \frac{\uvec L_i\cdot\uvec J - (\uvec L_i\cdot\uvec N_\infty)(\uvec J\cdot\uvec N_\infty)} {\uvec N_\infty\cdot(\uvec L_i\times\uvec J)} \,,\\
        \tan \Delta \psi^\rm{lens}_I &= \frac{\uvec N_\infty \cdot \left( \uvec L_i \times \uvec L_\rm{lens}^I \right)}{\uvec L_i \cdot \uvec L_\rm{lens}^I - \left( \uvec L_i \cdot \uvec N_\infty \right)\left( \uvec L_\rm{lens}^I \cdot \uvec N_\infty \right)} \,, \\
        \tan \Delta \psi^\rm{ab}_I &= \frac{\uvec N_\infty \cdot \left( \uvec L_\rm{lens}^I \times \uvec L_\rm{eff}^I \right)}{\uvec L_\rm{lens}^I \cdot \uvec L_\rm{eff}^I - \left( \uvec L_\rm{lens}^I \cdot \uvec N_\infty \right)\left( \uvec L_\rm{eff}^I \cdot \uvec N_\infty \right)} \,,
    \end{align}
\end{subequations}
where it is understood that we have suppressed the subscript $I$ for the polarization vector above. By introducing the quantities $\uvec L_\rm{lens}^I, \uvec L_\rm{eff}^I$, it concretely makes the equations simpler since we can cache the values of these numerically. Note that just as in the Thomas phase case, the flat-space polarization term does not explicitly depend on image $I$, but in the waveform its time dependence uses $\uvec L_i(t_o^I(\tau))$ implicitly.

In Fig.~\ref{fig:iota-psi-phiT-primary}, we show an example of how these effects combine for the primary image. In each of the panels, we plot the ray traced without bending/aberration (blue), ray traced with bending (green), and full ray traced model including bending and aberration (red). We show the locations that the system is maximally aligned (dashed grey) and when the outer orbit is at pericenter (dotted grey). This system has parameters $a_o = 50 M_3$, $e_o=0.3$, $\iota_J = 72^\circ$, and $\lambda_L = 25^\circ$, and we plot for $t_\rm{o} \in [0,3 P_r]$ on the x-axis. The flat LOS precesses throughout, which gives the gentle slope of the blue line in all three plots. The angular momentum precession rate slightly increases when the outer orbit is at pericenter which can be seen. If we look at the lensed LOS, we can see that lensing causes the inclination to be minimal where maximal alignment occurs, and additionally the polarization angle and Thomas phase change most here since LOS bending rotates most rapidly here. Finally, when including bending + aberration in the LOS, we can see that there is a peak in the inclination angle near pericenter (when velocity is largest). The polarization angle and Thomas phase now have more significant modulation due to aberration being more important in this regime. We only include the primary image in this plot, but for the secondary one these effects are magnified because the LOS of the secondary image is significantly different from the flat space one (since the bending angle of the secondary is larger than that of the primary). Figure \ref{fig:iota-psi-phiT-primary} demonstrates that the waveform can have significant modulations near lensing and aberration peaks which are important to correctly model.

\begin{figure*}[t]
    \centering
    \includegraphics{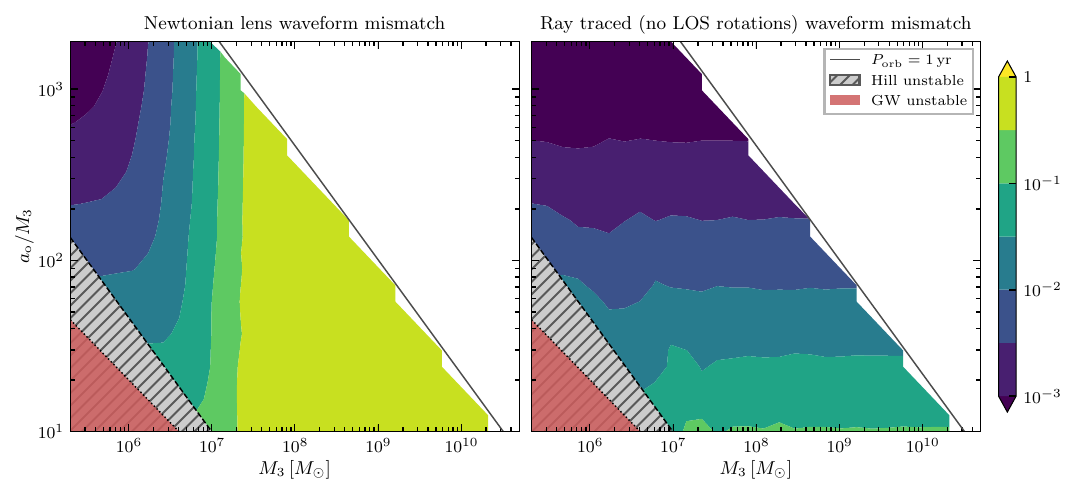}
    \caption{We show the astrophysical implications of these triple modulations by comparing the mismatch of the Newtonian lens waveform (left) and the ray-traced waveform without LOS rotations (right) to the one obtained in this manuscript. We choose outer orbit inclination $\iota_J = 60^\circ$, $f_\rm{gw} = 10^{-3} \,\rm{Hz}$, and  $e_o = 0$. One can see that the Newtonian lensing waveform on the left performs worse than the ray traced one where the waveform is completely misspecified for $M_3 \gtrsim 10^7 M_\odot$ because the time delay is incorrect ($f_\rm{gw}(\Delta t_\rm{exact} -\Delta t_\rm{0PN}) \ge 1$ here). On the right side, we show that the LOS rotations for the waveform give a mismatch scaling that goes as $M_3/a_o$. One can see that the Newtonian approach for triple waveforms is going to be inadequate for nearly the whole relevant parameter space, and rotations due to lensing and aberration are needed for correctly modeling the closest triples when $a_o/M_3 \lesssim 10^2$. We plot the region when the system is unstable to radiation reaction $E/\dot E \leq P_\rm{orb}$ for $m_1 + m_2 = 100\,M_\odot$ as done in \cite{Yu:2020dlm} (red); the hatched grey region marks inner binaries tidally disrupted by the SMBH, $a_\rm{in} > r_\rm{H}/2$ with $r_\rm{H}$ the Hill radius \cite{Hills:1988cvl}.}
    \label{fig:mismatch-M3-a-pair}
\end{figure*}

\subsection{Validation and mismatch}\label{app:validation-mismatch}

Let us validate when the effects that are computed in this paper are relevant for waveform modeling. In the main text, we compare to a `Newtonian lens' waveform which we define as
\begin{equation}\label{eq:newtonian-lens-waveform}
    h_\rm{new}(t) = \sum_I |\mu_I^\rm{pm}|^{1/2} \, h_\rm{ul}(t - \Delta t_I )
\end{equation}
where $\mu_I^\rm{pm}$ is the point-mass magnification from Eq.~\eqref{eq:magnification-geometric-0PN}, $\Delta t_1 = \Delta_\rm{0PN}$ is the Rømer time delay from Eq.~\eqref{eq:romer-time-delay-PN} for the primary image, and $ h_\rm{ul}(t - \Delta t_I )$ is the unlensed waveform. For the second image, we use the time delay $\Delta t_2 = \Delta_\rm{0PN} + \Delta t_{l}$, where $\Delta t_l$ is the time delay between images from Eq.~\eqref{eq:image-time-delay-0PN}. We compute the outer orbital motion and angular momentum precession using parallel transport, so the only difference is the lens and aberration model. The Newtonian lens waveform $h_\rm{new}(t)$ is the same as that shown in Fig.~\ref{fig:comparison-inc-85} of the main text and is a time-domain waveform that nearly matches some previous works\footnote{Note that these previous works don't exactly match because (1) they work in the frequency domain, (2) they evolve the inner binary's motion, and (3) they use PN approximations for de Sitter precession. With these assumptions pointed out, the mismatch between the `Newtonian lens' one and the one in our paper as presented in Fig.~\ref{fig:mismatch-M3-a-pair} would be extremely close.} \cite{Yu:2020dlm,Yu:2021dqx,Laeuger:2023qyz}.

Let us briefly define the mismatch between two waveforms, which characterizes how similar two templates are. The overlap $\mathcal{O}$ between two waveforms $h_1$ and $h_2$ is defined as 
\begin{equation}
    \mathcal{O}(h_1,h_2) = \max_{t_c,\phi_c} \frac{\left( h_1 | h_2(t_c,\phi_c) \right)}{\sqrt{\left( h_1 | h_1 \right) \left( h_2(t_c,\phi_c)|h_2(t_c,\phi_c) \right)}} \,,
\end{equation}
which is maximized over the time and phase of coalescence and the noise-weighted inner product is defined as 
\begin{equation}
    \left( h_1 | h_2 \right) \equiv 2 \int df \frac{\tilde h_1^* \tilde h_2 + \tilde h_2^* \tilde h_1}{S_n(f)}\,,
\end{equation}
where $S_n(f)$ is the power spectral density of the detector. In this manuscript, we use time-domain waveforms, and assume a flat power spectral density. For a flat PSD, the noise-weighted inner product can be expressed as 
\begin{equation}
    \left( h_1 | h_2 \right) = \frac{2}{S_n} \, \rm{Re} \int h_1^*(t)\, h_2(t)\, dt \,,
\end{equation}
where we use the complex time-domain strain $h = h_+ - i h_\times$ in this computation\footnote{Using the complex time-domain strain in the inner product like this is analogous to having two GW detectors which are tilted $45^\circ$ relative to one another.}. The mismatch between two waveforms can be expressed as 
\begin{equation}
    \mathcal{M}(h_1,h_2) = 1-\mathcal{O}(h_1,h_2)\,,
\end{equation}
which characterizes how similar these waveforms are. Note that we are doing this computation for a waveform which has multiple images, so the phase of coalescence $\phi_c$ is the phase in the source frame at merger. 
In Fig.~\ref{fig:mismatch-M3-a-pair}, we show the mismatch between our exact geometric-optics waveform and some reference triple waveforms. On the left panel, we plot the mismatch between the simple Newtonian lens waveform in Eq.~\eqref{eq:newtonian-lens-waveform} and the full geometric-optics result for a monochromatic GW signal with $\iota_J = 60^\circ$, $f_\rm{gw} = 10^{-3} \,\rm{Hz}$, and  $e_o = 0$. The mismatch is computed for one outer orbital cycle; since there is no merger, the phase of coalescence $\phi_c$ is taken to be the phase at $t_\rm{o} = 0$ instead. Since we are interested in systems that LISA can detect, we only compute the mismatch in regions where the outer orbit has a period of less than one year. We can see that the Newtonian lens waveform has mismatches which are $\mathcal{M} \ge 10^{-1}$ over much of the parameter space because the lensing time delay is not correct. On the right-hand side, we show the mismatch between the ray traced waveform with and without LOS rotations. Since the lensing time delays agree between these two systems, the error between these scales as $M_3/a_o$. Figure~\ref{fig:mismatch-M3-a-pair} demonstrates that for most of the parameter space, the errors from not ray tracing the time delays are dominant (left) compared to LOS rotations (right) for more astrophysically typical solutions with gravitational radii of $\sim10^2-10^3$. This manuscript demonstrates it is possible to get faithful GW hierarchical triple waveforms over the whole parameter space down to 10 gravitational radii, though this needs careful accounting of LOS rotations.

\bibliography{bibliography.bib}

\end{document}